\documentclass[%
preprint,
superscriptaddress,
 amsmath,amssymb,
 aps,
prapp,
]{revtex4-1}

\usepackage{graphicx}% Include figure files
\usepackage{dcolumn}% Align table columns on decimal point
\usepackage{bm}% bold math
\usepackage{braket}
\usepackage{siunitx}
\usepackage{hyperref}% add hypertext capabilities
\usepackage[mathlines]{lineno}% Enable numbering of text and display math

\begin{document}

%\preprint{APS/123-QED}

\title{Giant Photocurrent Enhancement by Coulomb Interaction in a Single Quantum Dot for Energy Harvesting}% Force line breaks with \\

\author{Kai Peng}
\author{Shiyao Wu}
\author{Xin Xie}
\author{Jingnan Yang}
\author{Chenjiang Qian}
\author{Feilong Song}
\author{Sibai Sun}
\author{Jianchen Dang}
\author{Yang Yu}
\author{Shushu Shi}
\author{Jiongji He}
\affiliation{Beijing National Laboratory for Condensed Matter Physics, Institute of Physics, Chinese Academy of Sciences, Beijing 100190, China}
\affiliation{CAS Center for Excellence in Topological Quantum Computation and School of Physical Sciences, University of Chinese Academy of Sciences, Beijing 100049, China}

\author{Xiulai Xu}%
\email{xlxu@iphy.ac.cn}
\affiliation{Beijing National Laboratory for Condensed Matter Physics, Institute of Physics, Chinese Academy of Sciences, Beijing 100190, China}
\affiliation{CAS Center for Excellence in Topological Quantum Computation and School of Physical Sciences, University of Chinese Academy of Sciences, Beijing 100049, China}
\affiliation{Songshan Lake Materials Laboratory, Dongguan, Guangdong 523808, China}

\date{\today}% It is always \today, today,
             %  but any date may be explicitly specified

\begin{abstract}

Understanding the carrier excitation and transport processes at the single-charge level plays a key role in quantum-dot-based solar cells and photodetectors. Here, we report on Coulomb-induced giant photocurrent enhancement of positive charged trions (\emph{X$^+$}) in a single self-assembled InAs/GaAs quantum dot embedded in an \emph{n-i-}Schottky device by high-resolution photocurrent (PC) spectroscopy.
The Coulomb repulsion between the two holes in the \emph{X$^+$} increases the tunneling rate of the hole, and the remaining hole can be reused as the initial state to regenerate \emph{X$^+$} again. This process brings the PC amplitude of \emph{X$^+$} up to 30 times larger than that of the neutral exciton.
%The enhancement is due to that the Coulomb repulsion between the two holes in the \emph{X$^+$} increases the tunneling rate of the hole, which brings the PC amplitude of \emph{X$^+$} to about 30 times larger than that of the normal neutral exciton {\emph{X$^0$}}.
The analysis of the hole tunneling time gives the equivalent change of hole tunnel barriers caused by Coulomb interaction between two holes with a value of 8.05 meV during the tunneling process.
%The hole tunneling time for single- and double-hole situations is obtained precisely according to a four-level rate-equation model through the pumping-power-dependent PC spectra, which gives the equivalent change of hole tunnel barriers caused by Coulomb interaction between two holes with a value of 9.65 meV during the tunneling process.
Our work brings a fundamental understanding of energy conversion for solar cells in nanoscale to improve internal quantum efficiency for energy harvesting.

\end{abstract}

\pacs{Valid PACS appear here}% PACS, the Physics and Astronomy
                             % Classification Scheme.
%\keywords{Suggested keywords}%Use showkeys class option if keyword
                              %display desired
\maketitle

%\tableofcontents
\section{\label{sec:level1}Introduction}

Semiconductor quantum dots (QDs) have attracted much attention as the third-generation photovoltaic solar cells due to the potential ultrahigh energy conversion efficiency \cite{green2006third,marti2003next}. Various approaches have been investigated intensively to improve the efficiency such as intermediate-band excitation utilizing low-energy photons \cite{luque1997increasing,tomi2010intermediate,luque2012understanding,okada2015intermediate}, enhancing electron transfer in sensitized solar cells \cite{diguna2007high,kamat2008quantum,kojima2009organometal,gonzalez2010modeling,kamat2013quantum,pan2014high}, or producing multiple excitons per single photon \cite{nozik2002quantum,schaller2004high,ellingson2005highly,schaller2006seven}. The dissociation of photogenerated excitons into free electrons and holes plays a key role in the solar cells and photodetectors \cite{scholes2006excitons}, which is affected by the strong Coulomb interactions between the carriers in the nanoscale systems.
Many researches have been focused on the mechanism of dissociation against the Coulomb attraction between electron and hole in the low-dimensional nanostructures \cite{ellingson2005highly,muntwiler2008coulomb,pijpers2009assessment,gelinas2013ultrafast,jailaubekov2013hot,jakowetz2017visualizing,blancon2017extremely}. While during the excitons' excitation and dissociation, various excitons which consist of different numbers of electrons and holes can be generated.
%The Coulomb repulsion between these carriers in the solar cells has an important influence on the photon absorption and carrier tunnelling, such as carrier multiplication in colloidal QDs \cite{pijpers2009assessment}, while it is hard to explore the strength of Coulomb interaction directly in experiments.
The effect of the Coulomb repulsion between these carriers in the solar cells has rarely been explored.
Recently, a strong enhancement of conversion efficiency with built-in electrons of intermediate-band QD solar cell has been reported \cite{kimberly2011strong}. Generally, the Coulomb repulsion between the electrons (holes) in the conduction (valance) band can accelerate the tunneling rate of electron (hole), while the mechanism still needs to be investigated quantitatively.
Photocurrent (PC) spectroscopy of a single QD at low temperature has been proved as a powerful method to investigate the hole-spin-based qubit in quantum information processing \cite{zrenner2002coherent,Ramsay2008Fast,mar2014ultrafast}, or the photon absorption and carrier tunneling process in the QD-based solar cells \cite{tomohiro2015direct}, which offers a profound understanding of the carriers dynamics in the applications of solar cell and photodetector based on QDs at the single-charge level.

In this paper, we demonstrate the Coulomb-induced giant enhancement of the PC in a single InAs/GaAs QD via high-resolution PC spectroscopy of positive charged trion (\emph{X$^+$}).
The QDs are embedded in the intrinsic region of an \emph{n-i-}Schottky photodiode based on a two-dimensional electron gas (2DEG). The two-color continuous wave (CW) narrow-bandwidth ($\sim$1MHz) lasers are used to perform the high-resolution PC spectra of \emph{X$^+$}, which are measured by sweeping the neutral exciton (\emph{X$^0$}) and \emph{X$^+$} transition energies simultaneously through quantum-confined Stark effect (QCSE) to achieve the resonant excitation. The Coulomb repulsion between the two holes in the \emph{X$^+$} increases the tunneling rate of one hole, and the remaining hole can be reused as the initial state to excite \emph{X$^+$} again. This process enhances the PC amplitude of \emph{X$^+$} dozens of times larger than that of {\emph{X$^0$}}.
The saturation behavior in the pumping-power-dependent PC measurements is intuitively interpreted by a four-level-rate-equation model, from which the hole tunneling time for \emph{X$^+$} and \emph{X$^0$} is obtained precisely.
By repeating the measurements for a range of excitation energies, we obtain the hole tunneling time as a function of vertical electric field. The Wentzel-Kramers-Brillouin (WKB) approximation is used to determine the tunnel barriers of hole in the s-shell of valence band for single- and double-hole situations quantitatively, which shows the change of 8.05 meV of the tunnel barrier caused by the hole repulsion. These results can be utilized to improve the photoelectric conversion efficiency and photoresponse in the applications of solar cells and photodetectors based on QDs.
\section{\label{sec:level1}EXPERIMENTAL DETAILS}

\begin{figure}
\includegraphics{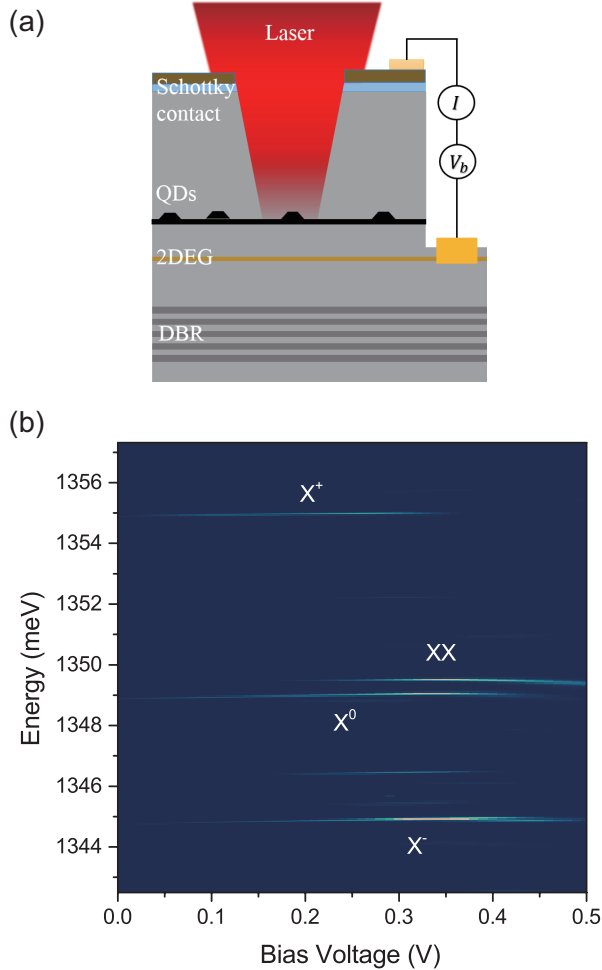}
\caption{\label{fig:1}The device structure and bias-dependent micro-PL spectra of a single QD. (a) Schematic diagram of the \emph{n-i-}Schottky photodiode based on 2DEG. (b) Bias-dependent micro-PL spectra of the single QD, showing PL emissions from \emph{X$^+$}, \emph{XX}, \emph{X$^0$} and \emph{X$^-$}.
}
\end{figure}

The \emph{n-i-}Schottky device is designed and fabricated for performing PC measurement of single QDs, where the device structure is shown schematically in Fig.~\ref{fig:1}(a). A single layer of InAs self-assemble QDs is grown by molecular beam epitaxy, which is embedded in a 250-nm-thick GaAs layer with a low density of about 10$^{9}$ cm$^{-2}$. A Si $\delta$-doped GaAs layer is located 50 nm below with a doping density $N_d$ = 5$\times$10$^{12}$ cm$^{-2}$ forming a 2DEG. The Schottky contact is formed by evaporating a 10 nm semitransparent Ti at the surface, followed by a Al mask with apertures of about 1-3 $\mu$m. A (Au, Ge)Ni ohmic contact is fabricated to connect the 2DEG with the Cr/Au bond pads. Moreover, a distributed Bragg reflector of 13-pairs of Al$_{0.94}$Ga$_{0.06}$As/GaAs (67/71 nm) is grown at the bottom of the structure to enhance the photon collection efficiency. The vertical electric field can be applied on the QDs as \emph{F} = (\emph{V$_i$} - \emph{V$_b$})/\emph{d}, where \emph{V$_i$}, \emph{V$_b$} and \emph{d} are built-in potential (0.74 V for this device), applied bias voltage and distance between the Schottky contact and 2DEG, respectively.

The device is placed on an xyz piezoelectric stage in the helium gas exchange cryostat at 4.2 K. A confocal microscopy with a large numerical aperture of NA = 0.82 microscope objective is used to perform micro-PL and PC measurements for single QDs. Nonresonant excitation is achieved by using a 650-nm semiconductor laser for PL measurement, and two tunable narrow-bandwidth ($\sim$1 MHz) external-cavity diode lasers in Littrow configuration are used to achieve resonant excitations. The PL signals of QDs are collected and dispersed through a 0.55-m spectrometer, and detected by a liquid-nitrogen-cooled charge coupled device camera with a spectra resolution of about 60 $\mu$eV. A semiconductor analyzer with a high current resolution (10 fA) is used to measure the current.

\section{\label{sec:level1}Results and discussion}

Before carrying out the PC measurements, bias-dependent micro-PL spectroscopy is performed on a single QD with above-band excitation to identify the transition energies of different charged exciton states and the bias voltage range for the PC regime, as shown in Fig.~\ref{fig:1}(b).
The \emph{$X^0$} and the biexciton (\emph{XX}) peaks have fine structure splitting caused by the electron-hole exchange interaction and structure asymmetry of the QD \cite{bayer2002fine}, and the \emph{$X^0$} PL intensity reaches saturation earlier than \emph{XX}. These behaviors help us to identify the \emph{$X^0$} and the \emph{XX} peaks. The charged trions can be identified by the binding energies and the electric-field-dependent behaviors.
At high positive bias voltages in Fig.~\ref{fig:1}(b), which correspond to low electric fields, the s-shell electron state is below the Fermi level in the 2DEG. The QDs will be charged with one electron tunneled from the 2DEG \cite{Mar2011bias}. As a result, the negative charged trion \emph{$X^-$} dominates.
While with the increase of the electric field, the s-shell electron level is above the Fermi level, and the electron charging stops. Instead, the tilt of the energy band makes the tunneling rate of captured electrons in the QDs faster than holes, resulting in the accumulation of holes in the QDs and the observation of positive charged trion \emph{$X^+$}. When the electric field is strong enough to let the hole tunnel out of the QD before the recombination with electron, the PL peaks of the QD disappear. At this regime, the PC can be observed.

\begin{figure*}
\includegraphics[scale=0.9]{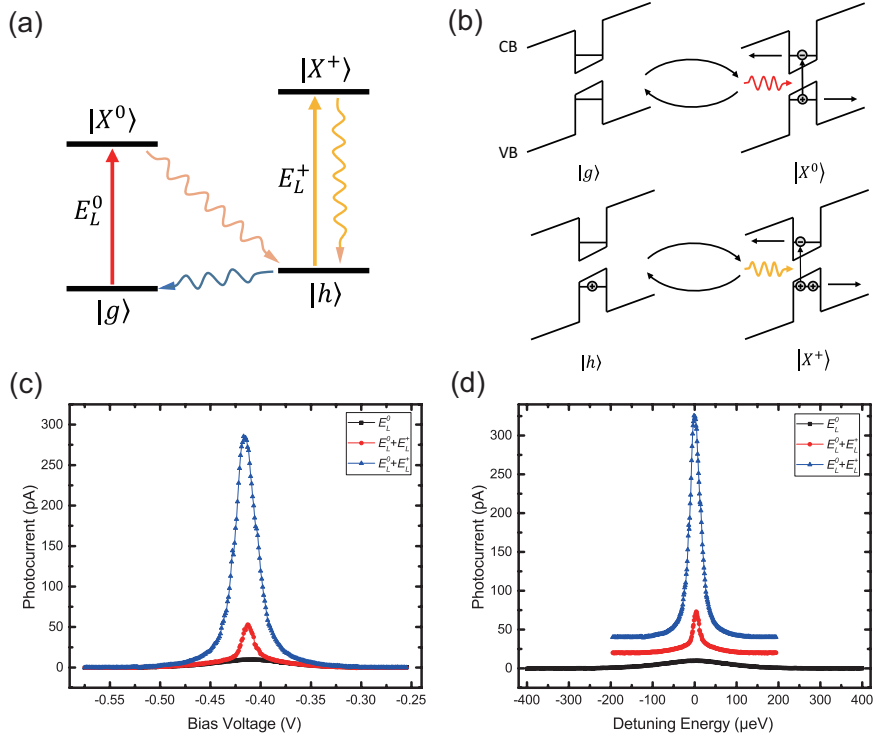}
\caption{\label{fig:2}The schematics of the PC measurements and the PC spectra of \emph{X$^0$} and \emph{X$^+$} from a single QD. (a) Two-color excitation scheme used for \emph{X$^+$} PC measurements. Here the solid arrows represent the excitation and the wavy arrows represent the tunneling processes of carriers. $\ket{g}$ represents the ground state of the QD, which can be excited to the neutral exciton state $\ket{X^0}$ resonantly through the first CW laser with energy \emph{E$^0_L$}.
The electron of \emph{X$^0$} tunnels out of the QD and a hole is left as $\ket{h}$. The second CW laser (\emph{E$^+_L$}) is used to excite the system to $\ket{X^+}$. Then the electron and one hole in \emph{$X^+$} will tunnel out of the dot and the system decays to single hole state $\ket{h}$. So the excitation from $\ket{h}$ to $\ket{X^+}$ can happen again. While the single hole may tunnel out of the dot and brings the system back to $\ket{g}$ for the next two-color excitation cycle. Two lasers are both linearly polarized, which assures that all the spin states can be excited for both \emph{X$^0$} and \emph{X$^+$}.
(b) Schematic illustration of PC for \emph{X$^0$} (top panel) and \emph{X$^+$} (bottom panel).
(c) PC spectra under the two-color excitation. The black solid squares represent the net \emph{X$^0$} PC spectrum with \emph{E$^0_L$} = 1347.68 meV; the red solid circles and the blue solid triangles represent the measured PC spectra under the two-color excitation with \emph{E$^+_L$} = 1354.05 meV at low and high excitation powers (50:1500), respectively. (d) The replotted PC spectra from Fig.~\ref{fig:2}(c) as a function of detuning energy. The bias voltage (electric filed) is replaced with energy through different QCSE for \emph{X$^0$} and \emph{X$^+$}. The Coulomb attraction of the extra hole to the electron induced linewidth narrowing is more clear. The spectra are shifted for clarity.
}
\end{figure*}

In the experiment, the PC spectra are measured by sweeping the exciton transition energy via QCSE to resonate with the fixed laser energy.
The Stark effect can be described as \emph{E(F)}=\emph{E}(0)+\emph{pF}+$\beta$\emph{F}$^2$, where \emph{E}(0) is the transition energy without external applied field, \emph{p} is the permanent dipole moment and $\beta$ is the polarizability of electron-hole wavefunctions. This is a convenient way to tune the transition energies of \emph{X$^0$} and \emph{X$^+$} simultaneously.
For \emph{$X^0$}, the energy band structures during the excitation and tunneling processes are shown as the top panel in Fig.~\ref{fig:2}(b). The QD s-shell is empty without laser shinning at negative bias voltages. The CW laser with energy $E^0_L$ can excite the QD from ground state $\ket{g}$ to $\ket{X^0}$ resonantly. Then the electron in the conduction band and the hole in the valance band will tunnel out of the dot under electric field, contributing to a measurable PC signal. As a result, the system is empty again and ready for the next excitation \cite{peng2017probing}.
While for \emph{X$^+$}, the exciton energy is renormalized due to the Coulomb interactions caused by the extra hole. In addition, the \emph{$X^+$} requires a single hole as the initial state, the two-color resonant excitation scheme is needed, as shown in Fig.~\ref{fig:2}(a).
Firstly, the \emph{X$^0$} is excited by a laser labeled as \emph{E$^0_L$} resonantly. Due to the fast tunneling rate of electron in the presence of electric field, the system will decay to the single hole state $\ket{h}$ in several picoseconds as the initial state of \emph{$X^+$}. Meanwhile, the second laser with higher energy (\emph{E$^+_L$}) pumps the QD to the $\ket{X^+}$ state. This is the two-color excitation scheme for \emph{X$^+$}.

There are different possible paths for the decay of \emph{$X^+$}. As mentioned above, the electron will tunnel out fast preferentially.
For the tunneling of holes, one hole tunnels out quickly due to the Coulomb repulsion between the two holes, and the system decays to $\ket{h}$. While the remaining hole decay to ground state $\ket{g}$ very slowly, which makes the $\ket{h}$ state as a metastable state to be excited to $\ket{X^+}$ again when the angular momentum condition is fulfilled, as the bottom panel shown in Fig.~\ref{fig:2}(b).
Therefore this $\ket{X^+}$$\rightarrow$$\ket{h}$$\rightarrow$$\ket{X^+}$ self-circulation process ensures that the excitation of \emph{X$^+$} does not totally depend on the $\ket{X^0}$$\rightarrow$$\ket{h}$ decay process. But it is still possible that the single hole tunnels out and the system returns to $\ket{g}$. Under this circumstance, the next two-color excitation loop can happen again. The net \emph{$X^+$} PC signal is from the $\ket{X^+}$$\rightarrow$$\ket{h}$ decay process, as shown in the bottom part of Fig.~\ref{fig:2}(b). Actually these two competitive decay paths coexist, that is why the two-color excitation is still needed although the circulation is already started.
It is worth noting that the spins of the carrier are ignored in the two-color excitation scheme. Here, the linearly polarized narrow-linewidth lasers are chosen to pump \emph{X$^0$} and \emph{X$^+$} resonantly, thus all the spin states can be excited for both \emph{X$^0$} and \emph{X$^+$} compared with the cross-circular-polarized scheme for spin-resolved excitation in previous works \cite{mar2013high,mar2014ultrafast}.

Figure ~\ref{fig:2}(c) shows the measured PC signals with the two-color excitation. Here, the energy of the first laser for exciting \emph{X$^0$} is fixed at \emph{E$^0_L$} = 1347.68 meV. The black solid points at the bottom represent the \emph{X$^0$} PC spectrum with only one-laser exciting. When the second laser with energy (\emph{E$^+_L$}) of 1354.05 meV is on, the \emph{X$^+$} PC components are added to the PC peak signal, as the red and blue curves shown in Fig.~\ref{fig:2}(c) for low and high pumping power (50:1500), respectively.
Fitted Lorentzian curve of the PC spectrum gives the corresponding central voltage on resonance, as well as the linewidth and the amplitude. The most striking feature of the two-color excitation PC signals is the giant enhancement of the PC amplitude, which is also observed for other QDs on similar Schottky devices in our experiments.
Here, the PC amplitude of \emph{X$^+$} is over one order of magnitude larger than that of \emph{X$^0$} at high excitation power surprisingly, while the previous works with cross-circular-polarized scheme for spin selection excitation can only achieve PC amplitude of \emph{X$^+$} comparable with that of \emph{X$^0$} \cite{mar2013high,mar2014ultrafast}.
One reason is that the reuse of hole from \emph{X$^+$} under linearly polarized excitation can remove the limit of the hole decay from \emph{X$^0$} partly. While for circularly polarized excitation scheme, the reuse of hole may not happen if this hole's transition needs perpendicular circularly polarized excitation.
More importantly, the Coulomb repulsion between the two holes increases the tunneling rate and enhances the PC amplitude largely.
It is also worth noting that a part of the PC signal shown as the blue points in Fig.~\ref{fig:2}(c) is from \emph{$X^0$}, while this component can be ignored. Because at high excitation power, the $\ket{X^+}$$\rightarrow$$\ket{h}$$\rightarrow$$\ket{X^+}$ self-circulation as shown in the bottom part in Fig.~\ref{fig:2}(b) dominates, the excitation of \emph{$X^0$} is restricted and its PC component is smaller than the net \emph{$X^0$} PC amplitude which is already over one order of magnitude smaller than that of \emph{$X^+$}.

\begin{figure}
\includegraphics[scale=0.85]{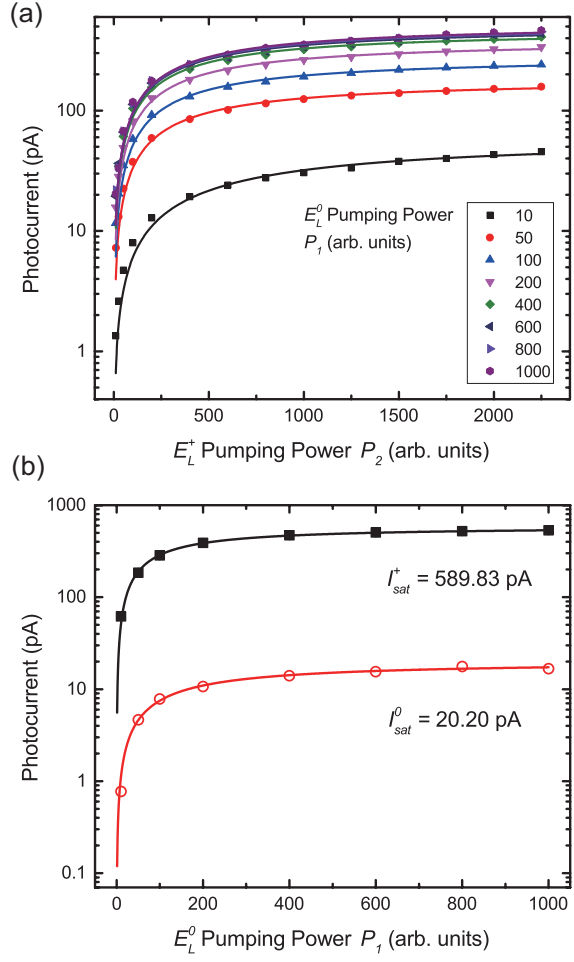}
\caption{\label{fig:3}Pumping-power-dependent PC amplitudes of \emph{X$^0$} and \emph{X$^+$}. (a) The \emph{$E^+_L$} pumping-power-dependent PC amplitude of \emph{X$^+$} for a series of \emph{$E^0_L$} pumping power $P_1$. (b) The \emph{$E^0_L$} pumping power-dependent PC amplitudes of \emph{X$^+$} and \emph{X$^0$}, corresponding to the solid squares and empty holes shown in the figure, respectively. The solid curves are the fitted results by using Eq.~(\ref{satur}).}
\end{figure}

To verify the Coulomb interactions between the extra hole in \emph{$X^+$} and other carriers, the linewidth of the PC spetrum is analyzed. The extra hole in \emph{$X^+$} increases the Coulomb attraction to the electron, which prolongs the tunneling time of electron and decreases the linewidth of the \emph{$X^+$} PC spectrum, as shown in Fig.~\ref{fig:2}(d) clearly, in which the x axis is replaced with the detuning energy through the different Stark effects of \emph{$X^0$} and \emph{$X^+$}. Both lasers are tuned simultaneously to obtain a series of PC spectra with different electric fields, and the correlations of transition energies and electric fields for \emph{$X^0$} and \emph{$X^+$} are built through quadratic fit according to QCSE.
Here, the linewidth of \emph{$X^+$} PC spectrum at low pumping power (red points) is about 20 $\mu$eV. If we ignore the power broadening at low pumping power and assume the electron tunneling as the main dephasing mechanism, the linewidth corresponds to the tunneling time of electron in \emph{$X^+$} of about 30 ps. As a contrast, the linewidth of \emph{$X^0$} PC spectrum is about 190 $\mu$eV in Fig.~\ref{fig:2}(d). Deducting the power broadening, the tunneling time of electron in \emph{$X^0$} is about several picoseconds, corresponding to the fast tunneling process from $\ket{X^0}$ to $\ket{h}$ shown in Fig.~\ref{fig:2}(a).

To demonstrate the giant enhancement of the PC amplitude under two-color excitation and the Coulomb interaction between two holes in \emph{X$^+$} quantitatively, the power-dependent PC measurements are performed. The saturation behavior in power-dependent measurement is a simple way to investigate the characteristic time of the system, which has been widely used in QD researches \cite{mar2011voltage,mar2011electrically,nguyen2012optically,bennett2016cavity,moody2016electronic,kurzmann2016auger}.
Here, the long tunneling time of hole limits the PC amplitudes for both \emph{X$^0$} and \emph{X$^+$}.
For the \emph{X$^0$} PC measurement, before the hole tunnels out of the QD, the next electron-hole pair cannot be excited resonantly by the same laser because of the energy detuning between different excitons. Therefore, a saturation of the PC amplitude of \emph{X$^0$} can be observed with the increase of the pumping power at a fixed electric field, as shown at the bottom of Fig.~\ref{fig:3}(b) (red line with empty holes). The saturation effect of the PC amplitude can be described by the following theoretical model \cite{beham2001nonlinear}:
\begin{equation}
I_{peak}=\frac{e}{2\tau^h_{esc}}\frac{\widetilde{P}}{\widetilde{P}+P_0}=I^0_{sat}\frac{\widetilde{P}}{\widetilde{P}+P_0},\label{satur}
\end{equation}
where $\widetilde{P}$ is the laser pumping power on the QD (arb. units), $P_0$ is the renomalized coefficient, \emph{e} is the elementary charge and $\tau^h_{esc}$ is the hole tunneling time. The power-dependent \emph{X$^0$} PC amplitudes can be fitted by this model very well, as the red curve shown in Fig.~\ref{fig:3}(b). Here, at \emph{F} = 46 kV/cm, the saturation PC amplitude \emph{$I^0_{sat}$} is 20.20 pA, which corresponds to the hole tunneling time as 3.96 ns.

But for \emph{X$^+$}, it is more complicated. The two-color excitation scheme makes the PC amplitude of \emph{X$^+$} depends on not only the pumping power of laser resonant with \emph{X$^+$}, but also the laser used to excite \emph{X$^0$}. Furthermore, the measured PC signals include the components from \emph{X$^+$} and \emph{X$^0$}, and the fast electron tunneling process from $\ket{X^0}$ to $\ket{h}$ also affects the preparation of \emph{X$^+$}. In order to describe the dynamics in the two-color excitation scheme shown in Fig.~\ref{fig:2}(a) precisely, we build the following four-level rate equations:
\begin{subequations}
\label{rate}
\begin{eqnarray}
&&\frac{dN_+}{dt}=-\frac{N_+}{\tau^+_{esc}}-B_{+h}N_+{\rho_2}+B_{h+}N_h{\rho_2},\label{equationa}
\\
&&\frac{dN_h}{dt}=\frac{N_+}{\tau^+_{esc}}+B_{+h}N_+{\rho_2}-B_{h+}N_h{\rho_2}-\frac{N_h}{\tau^h_{esc}}+\frac{N_x}{\tau^x_{esc}},\label{equationb}
\\
&&\frac{dN_x}{dt}=-B_{xg}N_x{\rho_1}-\frac{N_x}{\tau^x_{esc}}+B_{gx}N_g{\rho_1},\label{equationc}
\\
&&\frac{dN_g}{dt}=\frac{N_h}{\tau^h_{esc}}+B_{xg}N_x{\rho_1}-B_{gx}N_g{\rho_1}.\label{equationd}
\end{eqnarray}
\end{subequations}
Here, $N_+$, $N_h$, $N_x$, $N_g$ are the time-averaged occupation numbers of $\ket{X^+}$, $\ket{h}$, $\ket{X^0}$ and $\ket{g}$, respectively. $B_{+h}$, $B_{h+}$, $B_{xg}$, $B_{gx}$ are the stimulated emission and absorption coefficients, so we can set $B_{+h}$ = $B_{h+}$ = $B_2$ and $B_{xg}$ = $B_{gx}$ = $B_1$. $\rho_1$ and $\rho_2$ are the energy densities of the radiation field corresponding to the transitions of \emph{X$^0$} and \emph{X$^+$}, respectively, and they can be expressed in terms of the pumping power $P_1$ and $P_2$. The decay processes are introduced by $\tau^+_{esc}$, $\tau^h_{esc}$, and $\tau^x_{esc}$. Here, $\tau^+_{esc}$ and $\tau^h_{esc}$ correspond to the tunneling time of hole for double- and single-hole situations, respectively. The fast tunneling process from $\ket{X^0}$ to $\ket{h}$ is described by $\tau^x_{esc}$, which is the tunneling time of electron in \emph{X$^0$}. In this model, the spontaneous emissions are ignored in the electric field regime for PC measurement. The steady-state solutions can be obtained by considering the relation $N_+$ + $N_h$ + $N_x$ + $N_g$ = 1. The PC amplitude in the experiment can be described as:
\begin{equation}\label{ii}
  I=e\cdot(\frac{N_h}{\tau^h_{esc}}+\frac{N_+}{\tau^+_{esc}}),
\end{equation}
%where $\frac{N_h}{\tau^h_{esc}}$ and $\frac{N_+}{\tau^+_{esc}}$ correspond to the tunneling processes of \emph{X$^0$} and \emph{X$^+$} shown in Fig.~\ref{fig:2}(b), respectively.
where $N_h/\tau^h_{esc}$ and $N_+/\tau^+_{esc}$ correspond to the tunneling processes of \emph{X$^0$} and \emph{X$^+$} shown in Fig.~\ref{fig:2}(b), respectively.

Under the two-color excitation scheme, the PC amplitude depends on the pumping power of laser resonant with \emph{X$^+$} ($P_2$) and the laser used to excite \emph{X$^0$} ($P_1$). In order to get the hole tunneling time of \emph{X$^+$}, we perform two-step power-dependent measurements. At the beginning, we increase $P_2$ with a fixed $P_1$, so the saturation PC amplitude can be obtained through the saturation behavior which has the same form as Eq.~(\ref{satur}). Then a series of measurements for different $P_1$ are performed, as shown in Fig.~\ref{fig:3}(a), which give the saturation PC amplitudes with the increase of $P_1$, as shown by black line with solid squares in Fig.~\ref{fig:3}(b). Through the rate-equation model, the two-step power-dependent scheme gives the saturation PC amplitude in two-color excitation as:
\begin{equation}\label{jj}
  I_{sat}^+=\frac{e}{2\tau^+_{esc}}\cdot\frac{\tau^h_{esc}+\tau^+_{esc}}{\tau^h_{esc}+\tau^x_{esc}}\approx\frac{e}{2}\cdot(\frac{1}{\tau^+_{esc}}+\frac{1}{\tau^h_{esc}}).
\end{equation}
Here, the fast electron tunneling time $\tau^x_{esc}$ of about several picoseconds is ignored in Eq.~(\ref{jj}) compared with the hole tunneling time $\tau^h_{esc}$ of several nanoseconds. The saturation behaviors can be described very well with the model in Eq.~(\ref{jj}). Here, the saturation PC amplitude under two-color excitation is fitted as \emph{$I^+_{sat}$} = 589.83 pA, corresponding to the hole tunneling time for double-hole situation $\tau^+_{esc}$ as 0.14 ns, almost 30 times faster than that for single-hole situation as $\tau^h_{esc}$ = 3.96 ns.

It is not surprising that the saturation PC amplitude under two-color excitation depends on the tunneling rates of hole for single- and double-hole situations shown in Eq.~(\ref{jj}). Meanwhile the discrepancy up to 30 times of the tunneling rates proves that there is enough time for $\ket{h}$ to be excited to $\ket{X^+}$ mostly rather than decay to $\ket{g}$, as shown in Fig.~\ref{fig:2}(a).
The tunneling time of hole in \emph{$X^+$} (0.14 ns) is still much longer than that of electron ($\sim$30 ps), so the first hole tunneling process restricts the PC amplitude of \emph{$X^+$}.
Here at saturation, the $\ket{X^+}$$\rightarrow$$\ket{h}$$\rightarrow$$\ket{X^+}$ self-circulation process dominates the \emph{$X^+$} excitation and tunneling processes, and the hole tunneling time when the two holes occupy the valance band ground state limits the PC amplitude of \emph{$X^+$}.

\begin{figure}
\includegraphics[scale=0.8]{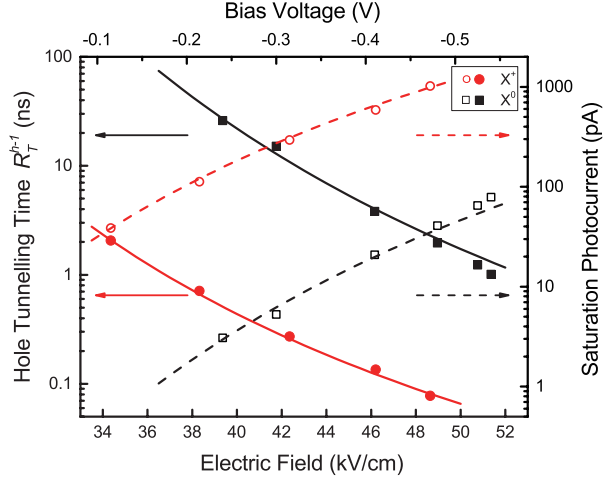}
\caption{\label{fig:4}Plot of hole tunneling time ${R^h_T}^{-1}$ and saturation PC $I_{peak}$ as a function of electric field (bias voltage). The solid and dashed lines are the fittings using the theoretical model in Eq.~(\ref{wkb}) based on 1D WKB approximation, yielding \emph{$E^0_b$}=45.51 meV and \emph{$E^+_b$}=37.46 meV with the QD height as 4.5 nm.}
\end{figure}

Obviously, the hole tunneling time $\tau^h_T$ is dependent on the electric field applied across the QD. We now repeat the power-dependent PC measurements of \emph{X$^0$} and \emph{X$^+$} for a series of electric fields by tuning the pumping laser energies of \emph{$E^0_L$} and \emph{$E^+_L$}, which give the saturation PC amplitude and hole tunneling time as a function of electric field for single- and double-hole situations, as shown in Fig.~\ref{fig:4}. The $\tau^h_T$ can be tuned more than one order of magnitude in the measured range of electric fields. For the QD, the hole tunneling rate \emph{$R^h_T$} (${\tau^h_T}^{-1}$) can be described by a one-dimensional (1D) (along the growth direction) WKB approximation:\cite{heller1998electric,oulton2002manipulation,tomohiro2015direct}
\begin{equation}
R^h_T=\frac{\hbar\pi}{2m^*_hH^2}\exp\lbrack\frac{-4}{3\hbar eF}\sqrt{2m^*_h E^3_b}\rbrack,\label{wkb}
\end{equation}
where $m^*_h$ = 0.59$m_e$ is the heavy-hole effective mass in GaAs along the growth direction and $m_e$ is the electron mass in vacuum, \emph{H} is the QD height, \emph{F} is the vertical electric field and $E_b$ is the tunnel barrier height or the ionization energy for the hole.
Even though the QDs have three-dimensional confinements, the confinement along the growth direction is much stronger than the QDs plane. As the high-resolution cross section image of a single QD from the same sample by transmission electron microscopy shown \cite{cao2015longitudinal}, the size of the QD can be measured as about 5 nm for height and about 20 nm for base length. On the other side, the electric filed is applied along the growth direction, so the 1D model is reasonable to describe the tunneling behavior in QDs along the growth direction.
This model of Eq.~(\ref{wkb}) can agree very well with the experimental data of electron and hole tunneling rates in the previous works \cite{mar2011electrically,mar2011voltage}.
As the device with the same structure, the QD height of 4.5 nm is chosen here to fit the experimental data shown in Fig.~\ref{fig:4}. The hole tunneling time and saturation PC of \emph{X$^0$}, as shown by black points in Fig.~\ref{fig:4}, give the fitted tunnel barrier \emph{$E^0_b$} of the single hole as 45.51 meV, which is consistent with the previous work \cite{mar2011electrically}.
Since the \emph{X$^+$} has two holes, the decay to single hole state has two tunneling channels, which induces the tunneling rate twice as fast as single hole tunneling without considering Coulomb interaction between two holes. So the hole tunneling rate for two-hole situation should be halved in Eq.~(\ref{wkb}) to fit the tunnel barrier \emph{$E^+_b$}.
%While for \emph{X$^+$} (the red points shown in the figure),
Here, the value of \emph{$E^+_b$} is 37.46 meV when two holes coexist. So equivalently, the Coulomb repulsion between the two holes provides about 8.05 meV change of the tunnel barrier, which describes the hole Coulomb repulsion interaction in the tunneling process quantitatively. And it is similar to the binding energy in PL of about 6 meV as shown in Fig.~\ref{fig:1}(b).

As shown in Fig.~\ref{fig:4}, the hole tunneling rate of \emph{$X^+$} is over one order of magnitude larger than that of \emph{$X^0$} in a large range of electric fields. Under a low electric field, the Coulomb interaction is more significant. Using the 1D WKB model in Eq.~(\ref{wkb}) with the fitted parameters, the extrapolated hole tunneling rate for double holes is more than two orders of magnitude larger than that for single hole at zero bias voltage. For the QDs researched, we can prepare \emph{$X^+$} and research the hole-hole Coulomb interaction due to the dominant tunneling of hole in the dissociation of excitons in a single QD through PC spectroscopy. We believe that this mechanism also works for the electron situation by redesigned device structure.
Although this QD device is not designed for solar cells particularly, the results however strongly prove that the Coulomb-induced giant enhancement of PC signal can work for the application of QD-based solar cells (with zero bias voltage), especially for the right QDs size for Coulomb energy and suitable barrier height.
For the practical applications for QD-based solar cells, the \emph{p-} or \emph{n-}doping QDs can be achieved with growing a two-dimensional hole- or electron-gas layer close to the QDs layer to prepare positive or negative charged trions, respectively. For example, a 50$\%$ increase of conversion efficiency in intermediate-band QDs solar cells with \emph{n-}doping has been reported \cite{kimberly2011strong}, where the interelectron Coulomb interaction can transfer electrons to the conducting state.
Furthermore, we believe that this mechanism of Coulomb-induced enhancement of tunneling can be extended to colloidal-QD- and perovskite-nanocrystal-based solar cells, even though the strength of Coulomb interactions between carriers might be different.

%For the QDs we researched, the hole tunneling process dominates the dissociation of the exciton. Due to that, the \emph{X$^+$} can be excited with two-color excitation
%For the practical applications for QD-based solar cells, the p- or n-doping QDs can be grown to create positive or negative charged trions.
%A 50$\%$ increase of conversion efficiency in InAs/GaAs QDs solar cells with n-doping has been reported \cite{kimberly2011strong}, where the interelectron Coulomb interaction can transfer electrons to the conducting state. For the QDs device we researched, the hole tunneling dominates the dissociation of excitons.

\section{\label{sec:level1}Conclusion}

In conclusion, we demonstrate the Coulomb-induced giant enhancement of \emph{X$^+$} PC in a single InAs/GaAs QD under the two-color excitation scheme. The high-resolution PC spectra of \emph{X$^+$} are obtained by sweeping the \emph{X$^0$} and \emph{X$^+$} transition energies to match the fixed narrow-bandwidth lasers via QCSE simultaneously. The Coulomb repulsion between the two holes in \emph{X$^+$}  greatly enhances the tunneling rate of hole, and the remaining hole can be reused to build the $\ket{X^+}$$\rightarrow$$\ket{h}$$\rightarrow$$\ket{X^+}$ self-circulation process under linearly polarized excitation scheme. This process increases the PC amplitude of \emph{X$^+$} by up to 30 times larger than that of \emph{X$^0$}. The hole tunneling time of \emph{X$^+$} and \emph{X$^0$} are successfully extracted from the saturation PC amplitude in the pumping-power-dependent PC spectra according to a four-level rate-equation model. The hole tunneling time as a function of electric field is achieved through performing the measurements for a range of bias voltages. By using the 1D WKB approximation model with a reasonable QD height, the tunnel barrier heights are fitted for single- and double-hole situations. The results show that the Coulomb repulsion offers about 8.05 meV change of the tunnel barrier, which enhances the hole tunneling rate greatly. This quantitative investigation of the Coulomb interactions in the few particle states and the giant enhancement of the PC amplitude in a single QD with only a single extra charge can have great potential applications for the hole-spin-based quantum information processing, also provide a new method to enhance the conversion efficiency for energy harvesting in solar cells and photodetectors based on semiconductor QDs.

\begin{acknowledgments}
This work was supported by the National Natural Science Foundation of China under Grants No. 61675228, No. 11721404, No. 51761145104 and No. 11874419; the Strategic Priority Research Program, the Instrument Developing Project and the Interdisciplinary Innovation Team of the Chinese Academy of Sciences under Grants No. XDB07030200, No. XDB28000000 and No.YJKYYQ20180036.
\end{acknowledgments}

%

%\appendix

% The \nocite command causes all entries in a bibliography to be printed out
% whether or not they are actually referenced in the text. This is appropriate
% for the sample file to show the different styles of references, but authors
% most likely will not want to use it.
\nocite{*}

%\bibliography{apssamp}% Produces the bibliography via BibTeX.

\begin{thebibliography}{40}%
\makeatletter
\providecommand \@ifxundefined [1]{%
 \@ifx{#1\undefined}
}%
\providecommand \@ifnum [1]{%
 \ifnum #1\expandafter \@firstoftwo
 \else \expandafter \@secondoftwo
 \fi
}%
\providecommand \@ifx [1]{%
 \ifx #1\expandafter \@firstoftwo
 \else \expandafter \@secondoftwo
 \fi
}%
\providecommand \natexlab [1]{#1}%
%\providecommand \enquote  [1]{``#1''}%
\providecommand \enquote  [1]{#1}%
\providecommand \bibnamefont  [1]{#1}%
\providecommand \bibfnamefont [1]{#1}%
\providecommand \citenamefont [1]{#1}%
\providecommand \href@noop [0]{\@secondoftwo}%
\providecommand \href [0]{\begingroup \@sanitize@url \@href}%
\providecommand \@href[1]{\@@startlink{#1}\@@href}%
\providecommand \@@href[1]{\endgroup#1\@@endlink}%
\providecommand \@sanitize@url [0]{\catcode `\\12\catcode `\$12\catcode
  `\&12\catcode `\#12\catcode `\^12\catcode `\_12\catcode `\%12\relax}%
\providecommand \@@startlink[1]{}%
\providecommand \@@endlink[0]{}%
\providecommand \url  [0]{\begingroup\@sanitize@url \@url }%
\providecommand \@url [1]{\endgroup\@href {#1}{\urlprefix }}%
\providecommand \urlprefix  [0]{URL }%
\providecommand \Eprint [0]{\href }%
\providecommand \doibase [0]{http://dx.doi.org/}%
\providecommand \selectlanguage [0]{\@gobble}%
\providecommand \bibinfo  [0]{\@secondoftwo}%
\providecommand \bibfield  [0]{\@secondoftwo}%
\providecommand \translation [1]{[#1]}%
\providecommand \BibitemOpen [0]{}%
\providecommand \bibitemStop [0]{}%
\providecommand \bibitemNoStop [0]{.\EOS\space}%
\providecommand \EOS [0]{\spacefactor3000\relax}%
\providecommand \BibitemShut  [1]{\csname bibitem#1\endcsname}%
\let\auto@bib@innerbib\@empty
%</preamble>
\bibitem [{\citenamefont {Green}(2006)}]{green2006third}%
  \BibitemOpen
  \bibfield  {author} {\bibinfo {author} {\bibfnamefont {M.~A.}\ \bibnamefont
  {Green}},\ }\href@noop {} {\emph {\bibinfo {title} {Third generation
  photovoltaics}}}\ (\bibinfo  {publisher} {Springer},\ \bibinfo {year}
  {2006})\BibitemShut {NoStop}%
\bibitem [{\citenamefont {Mart{\'\i}}\ and\ \citenamefont
  {Luque}(2003)}]{marti2003next}%
  \BibitemOpen
  \bibfield  {author} {\bibinfo {author} {\bibfnamefont {A.}~\bibnamefont
  {Mart{\'\i}}}\ and\ \bibinfo {author} {\bibfnamefont {A.}~\bibnamefont
  {Luque}},\ }\href@noop {} {\emph {\bibinfo {title} {Next generation
  photovoltaics: high efficiency through full spectrum utilization}}}\
  (\bibinfo  {publisher} {Institute of Physics},\ \bibinfo {year}
  {2003})\BibitemShut {NoStop}%
\bibitem [{\citenamefont {Luque}\ and\ \citenamefont
  {Mart\'{\i}}(1997)}]{luque1997increasing}%
  \BibitemOpen
  \bibfield  {author} {\bibinfo {author} {\bibfnamefont {A.}~\bibnamefont
  {Luque}}\ and\ \bibinfo {author} {\bibfnamefont {A.}~\bibnamefont
  {Mart\'{\i}}},\ }\bibfield  {title} {\enquote {\bibinfo {title} {Increasing the Efficiency of Ideal Solar Cells by Photon Induced Transitions at Intermediate Levels},}\ }\href {\doibase 10.1103/PhysRevLett.78.5014} {\bibfield
  {journal} {\bibinfo  {journal} {Phys. Rev. Lett.}\ }\textbf {\bibinfo
  {volume} {78}},\ \bibinfo {pages} {5014} (\bibinfo {year}
  {1997})}\BibitemShut {NoStop}%
\bibitem [{\citenamefont {Tomi\ifmmode~\acute{c}\else
  \'{c}\fi{}}(2010)}]{tomi2010intermediate}%
  \BibitemOpen
  \bibfield  {author} {\bibinfo {author} {\bibfnamefont {S.}~\bibnamefont
  {Tomi\ifmmode~\acute{c}\else \'{c}\fi{}}},\ }\bibfield  {title} {\enquote {\bibinfo {title} {Intermediate-band solar cells: Influence of band formation on dynamical processes in InAs/GaAs quantum dot arrays},}\ }\href {\doibase
  10.1103/PhysRevB.82.195321} {\bibfield  {journal} {\bibinfo  {journal} {Phys.
  Rev. B}\ }\textbf {\bibinfo {volume} {82}},\ \bibinfo {pages} {195321}
  (\bibinfo {year} {2010})}\BibitemShut {NoStop}%
\bibitem [{\citenamefont {Luque}\ \emph {et~al.}(2012)\citenamefont {Luque},
  \citenamefont {Mart{\'\i}},\ and\ \citenamefont
  {Stanley}}]{luque2012understanding}%
  \BibitemOpen
  \bibfield  {author} {\bibinfo {author} {\bibfnamefont {A.}~\bibnamefont
  {Luque}}, \bibinfo {author} {\bibfnamefont {A.}~\bibnamefont {Mart{\'\i}}}, \
  and\ \bibinfo {author} {\bibfnamefont {C.}~\bibnamefont {Stanley}},\ }\bibfield  {title} {\enquote {\bibinfo {title} {Understanding intermediate-band solar cells},}\ }\href
  {https://www.nature.com/articles/nphoton.2012.1} {\bibfield  {journal}
  {\bibinfo  {journal} {Nat. Photonics}\ }\textbf {\bibinfo {volume} {6}},\
  \bibinfo {pages} {146} (\bibinfo {year} {2012})}\BibitemShut {NoStop}%
\bibitem [{\citenamefont {Okada}\ \emph {et~al.}(2015)\citenamefont {Okada},
  \citenamefont {Ekins-Daukes}, \citenamefont {Kita}, \citenamefont {Tamaki},
  \citenamefont {Yoshida}, \citenamefont {Pusch}, \citenamefont {Hess},
  \citenamefont {Phillips}, \citenamefont {Farrell}, \citenamefont {Yoshida},
  \citenamefont {Ahsan}, \citenamefont {Shoji}, \citenamefont {Sogabe},\ and\
  \citenamefont {Guillemoles}}]{okada2015intermediate}%
  \BibitemOpen
  \bibfield  {author} {\bibinfo {author} {\bibfnamefont {Y.}~\bibnamefont
  {Okada}}, \bibinfo {author} {\bibfnamefont {N.~J.}\ \bibnamefont
  {Ekins-Daukes}}, \bibinfo {author} {\bibfnamefont {T.}~\bibnamefont {Kita}},
  \bibinfo {author} {\bibfnamefont {R.}~\bibnamefont {Tamaki}}, \bibinfo
  {author} {\bibfnamefont {M.}~\bibnamefont {Yoshida}}, \bibinfo {author}
  {\bibfnamefont {A.}~\bibnamefont {Pusch}}, \bibinfo {author} {\bibfnamefont
  {O.}~\bibnamefont {Hess}}, \bibinfo {author} {\bibfnamefont {C.~C.}\
  \bibnamefont {Phillips}}, \bibinfo {author} {\bibfnamefont {D.~J.}\
  \bibnamefont {Farrell}}, \bibinfo {author} {\bibfnamefont {K.}~\bibnamefont
  {Yoshida}}, \bibinfo {author} {\bibfnamefont {N.}~\bibnamefont {Ahsan}},
  \bibinfo {author} {\bibfnamefont {Y.}~\bibnamefont {Shoji}}, \bibinfo
  {author} {\bibfnamefont {T.}~\bibnamefont {Sogabe}}, \ and\ \bibinfo {author}
  {\bibfnamefont {J.-F.}\ \bibnamefont {Guillemoles}},\ }\bibfield  {title} {\enquote {\bibinfo {title} {Intermediate band solar cells: Recent progress and future directions},}\ }\href {\doibase
  10.1063/1.4916561} {\bibfield  {journal} {\bibinfo  {journal} {Appl. Phys.
  Rev}\ }\textbf {\bibinfo {volume} {2}},\ \bibinfo {pages} {021302} (\bibinfo
  {year} {2015})}\BibitemShut {NoStop}%
\bibitem [{\citenamefont {Diguna}\ \emph {et~al.}(2007)\citenamefont {Diguna},
  \citenamefont {Shen}, \citenamefont {Kobayashi},\ and\ \citenamefont
  {Toyoda}}]{diguna2007high}%
  \BibitemOpen
  \bibfield  {author} {\bibinfo {author} {\bibfnamefont {L.~J.}\ \bibnamefont
  {Diguna}}, \bibinfo {author} {\bibfnamefont {Q.}~\bibnamefont {Shen}},
  \bibinfo {author} {\bibfnamefont {J.}~\bibnamefont {Kobayashi}}, \ and\
  \bibinfo {author} {\bibfnamefont {T.}~\bibnamefont {Toyoda}},\ }\bibfield  {title} {\enquote {\bibinfo {title} {High efficiency of CdSe quantum-dot-sensitized TiO2 inverse opal solar cells},}\ }\href
  {\doibase 10.1063/1.2757130} {\bibfield  {journal} {\bibinfo  {journal}
  {Appl. Phys. Lett}\ }\textbf {\bibinfo {volume} {91}},\ \bibinfo {pages}
  {023116} (\bibinfo {year} {2007})}\BibitemShut {NoStop}%
\bibitem [{\citenamefont {Kamat}(2008)}]{kamat2008quantum}%
  \BibitemOpen
  \bibfield  {author} {\bibinfo {author} {\bibfnamefont {P.~V.}\ \bibnamefont
  {Kamat}},\ }\bibfield  {title} {\enquote {\bibinfo {title} {Quantum Dot Solar Cells. Semiconductor Nanocrystals as Light Harvesters},}\ }\href {\doibase 10.1021/jp806791s} {\bibfield  {journal}
  {\bibinfo  {journal} {J. Phys. Chem. C}\ }\textbf {\bibinfo {volume} {112}},\
  \bibinfo {pages} {18737} (\bibinfo {year} {2008})}\BibitemShut {NoStop}%
\bibitem [{\citenamefont {Kojima}\ \emph {et~al.}(2009)\citenamefont {Kojima},
  \citenamefont {Teshima}, \citenamefont {Shirai},\ and\ \citenamefont
  {Miyasaka}}]{kojima2009organometal}%
  \BibitemOpen
  \bibfield  {author} {\bibinfo {author} {\bibfnamefont {A.}~\bibnamefont
  {Kojima}}, \bibinfo {author} {\bibfnamefont {K.}~\bibnamefont {Teshima}},
  \bibinfo {author} {\bibfnamefont {Y.}~\bibnamefont {Shirai}}, \ and\ \bibinfo
  {author} {\bibfnamefont {T.}~\bibnamefont {Miyasaka}},\ }\bibfield  {title} {\enquote {\bibinfo {title} {Organometal halide perovskites as visible-light sensitizers for photovoltaic Cells},}\ }\href {\doibase
  10.1021/ja809598r} {\bibfield  {journal} {\bibinfo  {journal} {J. Am. Chem.
  Soc}\ }\textbf {\bibinfo {volume} {131}},\ \bibinfo {pages} {6050} (\bibinfo
  {year} {2009})}\BibitemShut {NoStop}%
\bibitem [{\citenamefont {Gonz치lez-Pedro}\ \emph {et~al.}(2010)\citenamefont
  {Gonz치lez-Pedro}, \citenamefont {Xu}, \citenamefont {Mora-Ser 칩},\ and\
  \citenamefont {Bisquert}}]{gonzalez2010modeling}%
  \BibitemOpen
  \bibfield  {author} {\bibinfo {author} {\bibfnamefont {V.}~\bibnamefont
  {Gonz치lez-Pedro}}, \bibinfo {author} {\bibfnamefont {X.}~\bibnamefont {Xu}},
  \bibinfo {author} {\bibfnamefont {I.}~\bibnamefont {Mora-Ser 칩}}, \ and\
  \bibinfo {author} {\bibfnamefont {J.}~\bibnamefont {Bisquert}},\ }\bibfield  {title} {\enquote {\bibinfo {title} {Modeling high-efficiency quantum dot sensitized solar cells},}\ }\href
  {\doibase 10.1021/nn101534y} {\bibfield  {journal} {\bibinfo  {journal} {ACS
  Nano}\ }\textbf {\bibinfo {volume} {4}},\ \bibinfo {pages} {5783} (\bibinfo
  {year} {2010})}\BibitemShut {NoStop}%
\bibitem [{\citenamefont {Kamat}(2013)}]{kamat2013quantum}%
  \BibitemOpen
  \bibfield  {author} {\bibinfo {author} {\bibfnamefont {P.~V.}\ \bibnamefont
  {Kamat}},\ }\bibfield  {title} {\enquote {\bibinfo {title} {Quantum dot solar cells. the next big thing in photovoltaics},}\ }\href {\doibase 10.1021/jz400052e} {\bibfield  {journal}
  {\bibinfo  {journal} {J. Phys. Chem. C}\ }\textbf {\bibinfo {volume} {4}},\
  \bibinfo {pages} {908} (\bibinfo {year} {2013})}\BibitemShut {NoStop}%
\bibitem [{\citenamefont {Pan}\ \emph {et~al.}(2014)\citenamefont {Pan},
  \citenamefont {Mora-Ser\ifmmode~\acute{o}\else \'{o}\fi{}}, \citenamefont
  {Shen}, \citenamefont {Zhang}, \citenamefont {Li}, \citenamefont {Zhao},
  \citenamefont {Wang}, \citenamefont {Zhong},\ and\ \citenamefont
  {Bisquert}}]{pan2014high}%
  \BibitemOpen
  \bibfield  {author} {\bibinfo {author} {\bibfnamefont {Z.}~\bibnamefont
  {Pan}}, \bibinfo {author} {\bibfnamefont {I.}~\bibnamefont
  {Mora-Ser\ifmmode~\acute{o}\else \'{o}\fi{}}}, \bibinfo {author}
  {\bibfnamefont {Q.}~\bibnamefont {Shen}}, \bibinfo {author} {\bibfnamefont
  {H.}~\bibnamefont {Zhang}}, \bibinfo {author} {\bibfnamefont
  {Y.}~\bibnamefont {Li}}, \bibinfo {author} {\bibfnamefont {K.}~\bibnamefont
  {Zhao}}, \bibinfo {author} {\bibfnamefont {J.}~\bibnamefont {Wang}}, \bibinfo
  {author} {\bibfnamefont {X.}~\bibnamefont {Zhong}}, \ and\ \bibinfo {author}
  {\bibfnamefont {J.}~\bibnamefont {Bisquert}},\ }\bibfield  {title} {\enquote {\bibinfo {title} {High-efficiency 몷green몸 quantum dot solar cells},}\ }\href
  {https://pubs.acs.org/doi/abs/10.1021/ja504310w} {\bibfield  {journal}
  {\bibinfo  {journal} {J. Am. Chem. Soc}\ }\textbf {\bibinfo {volume} {136}},\
  \bibinfo {pages} {9203} (\bibinfo {year} {2014})}\BibitemShut {NoStop}%
\bibitem [{\citenamefont {Nozik}(2002)}]{nozik2002quantum}%
  \BibitemOpen
  \bibfield  {author} {\bibinfo {author} {\bibfnamefont {A.}~\bibnamefont
  {Nozik}},\ }\bibfield  {title} {\enquote {\bibinfo {title} {Quantum dot solar cells},}\ }\bibfield  {title} {\enquote {\bibinfo {title} {Quantum dot solar cells},}\ }\href
  {https://www.sciencedirect.com/science/article/pii/S1386947702003740}
  {\bibfield  {journal} {\bibinfo  {journal} {Physica E}\ }\textbf {\bibinfo
  {volume} {14}},\ \bibinfo {pages} {115} (\bibinfo {year} {2002})}\BibitemShut
  {NoStop}%
\bibitem [{\citenamefont {Schaller}\ and\ \citenamefont
  {Klimov}(2004)}]{schaller2004high}%
  \BibitemOpen
  \bibfield  {author} {\bibinfo {author} {\bibfnamefont {R.~D.}\ \bibnamefont
  {Schaller}}\ and\ \bibinfo {author} {\bibfnamefont {V.~I.}\ \bibnamefont
  {Klimov}},\ }\bibfield  {title} {\enquote {\bibinfo {title} {High Efficiency Carrier Multiplication in PbSe Nanocrystals: Implications for Solar Energy Conversion},}\ }\href {\doibase 10.1103/PhysRevLett.92.186601} {\bibfield
  {journal} {\bibinfo  {journal} {Phys. Rev. Lett.}\ }\textbf {\bibinfo
  {volume} {92}},\ \bibinfo {pages} {186601} (\bibinfo {year}
  {2004})}\BibitemShut {NoStop}%
\bibitem [{\citenamefont {Ellingson}\ \emph {et~al.}(2005)\citenamefont
  {Ellingson}, \citenamefont {Beard}, \citenamefont {Johnson}, \citenamefont
  {Yu}, \citenamefont {Micic}, \citenamefont {Nozik}, \citenamefont {Shabaev},\
  and\ \citenamefont {Efros}}]{ellingson2005highly}%
  \BibitemOpen
  \bibfield  {author} {\bibinfo {author} {\bibfnamefont {R.~J.}\ \bibnamefont
  {Ellingson}}, \bibinfo {author} {\bibfnamefont {M.~C.}\ \bibnamefont
  {Beard}}, \bibinfo {author} {\bibfnamefont {J.~C.}\ \bibnamefont {Johnson}},
  \bibinfo {author} {\bibfnamefont {P.}~\bibnamefont {Yu}}, \bibinfo {author}
  {\bibfnamefont {O.~I.}\ \bibnamefont {Micic}}, \bibinfo {author}
  {\bibfnamefont {A.~J.}\ \bibnamefont {Nozik}}, \bibinfo {author}
  {\bibfnamefont {A.}~\bibnamefont {Shabaev}}, \ and\ \bibinfo {author}
  {\bibfnamefont {A.~L.}\ \bibnamefont {Efros}},\ }\bibfield  {title} {\enquote {\bibinfo {title} {Highly efficient multiple exciton generation in colloidal PbSe and PbS quantum dots},}\ }\href {\doibase
  10.1021/nl0502672} {\bibfield  {journal} {\bibinfo  {journal} {Nano Lett}\
  }\textbf {\bibinfo {volume} {5}},\ \bibinfo {pages} {865} (\bibinfo {year}
  {2005})}\BibitemShut {NoStop}%
\bibitem [{\citenamefont {Schaller}\ \emph {et~al.}(2006)\citenamefont
  {Schaller}, \citenamefont {Sykora}, \citenamefont {Pietryga},\ and\
  \citenamefont {Klimov}}]{schaller2006seven}%
  \BibitemOpen
  \bibfield  {author} {\bibinfo {author} {\bibfnamefont {R.~D.}\ \bibnamefont
  {Schaller}}, \bibinfo {author} {\bibfnamefont {M.}~\bibnamefont {Sykora}},
  \bibinfo {author} {\bibfnamefont {J.~M.}\ \bibnamefont {Pietryga}}, \ and\
  \bibinfo {author} {\bibfnamefont {V.~I.}\ \bibnamefont {Klimov}},\ }\bibfield  {title} {\enquote {\bibinfo {title} {Seven excitons at a cost of one: redefining the limits for conversion efficiency of photons into charge carriers},}\ }\href
  {https://pubs.acs.org/doi/abs/10.1021/nl052276g} {\bibfield  {journal}
  {\bibinfo  {journal} {Nano Lett}\ }\textbf {\bibinfo {volume} {6}},\ \bibinfo
  {pages} {424} (\bibinfo {year} {2006})}\BibitemShut {NoStop}%
\bibitem [{\citenamefont {Scholes}\ and\ \citenamefont
  {Rumbles}(2006)}]{scholes2006excitons}%
  \BibitemOpen
  \bibfield  {author} {\bibinfo {author} {\bibfnamefont {G.~D.}\ \bibnamefont
  {Scholes}}\ and\ \bibinfo {author} {\bibfnamefont {G.}~\bibnamefont
  {Rumbles}},\ }\bibfield  {title} {\enquote {\bibinfo {title} {Excitons in nanoscale systems},}\ }\href {https://www.nature.com/articles/nmat1710} {\bibfield
  {journal} {\bibinfo  {journal} {Nat. Mater}\ }\textbf {\bibinfo {volume}
  {5}},\ \bibinfo {pages} {683} (\bibinfo {year} {2006})}\BibitemShut {NoStop}%
\bibitem [{\citenamefont {Muntwiler}\ \emph {et~al.}(2008)\citenamefont
  {Muntwiler}, \citenamefont {Yang}, \citenamefont {Tisdale},\ and\
  \citenamefont {Zhu}}]{muntwiler2008coulomb}%
  \BibitemOpen
  \bibfield  {author} {\bibinfo {author} {\bibfnamefont {M.}~\bibnamefont
  {Muntwiler}}, \bibinfo {author} {\bibfnamefont {Q.}~\bibnamefont {Yang}},
  \bibinfo {author} {\bibfnamefont {W.~A.}\ \bibnamefont {Tisdale}}, \ and\
  \bibinfo {author} {\bibfnamefont {X.-Y.}\ \bibnamefont {Zhu}},\ }\bibfield  {title} {\enquote {\bibinfo {title} {Coulomb Barrier for Charge Separation at an Organic Semiconductor Interface},}\ }\href
  {\doibase 10.1103/PhysRevLett.101.196403} {\bibfield  {journal} {\bibinfo
  {journal} {Phys. Rev. Lett.}\ }\textbf {\bibinfo {volume} {101}},\ \bibinfo
  {pages} {196403} (\bibinfo {year} {2008})}\BibitemShut {NoStop}%
\bibitem [{\citenamefont {Pijpers}\ \emph {et~al.}(2009)\citenamefont
  {Pijpers}, \citenamefont {Ulbricht}, \citenamefont {Tielrooij}, \citenamefont
  {Osherov}, \citenamefont {Golan}, \citenamefont {Delerue}, \citenamefont
  {Allan},\ and\ \citenamefont {Bonn}}]{pijpers2009assessment}%
  \BibitemOpen
  \bibfield  {author} {\bibinfo {author} {\bibfnamefont {J.}~\bibnamefont
  {Pijpers}}, \bibinfo {author} {\bibfnamefont {R.}~\bibnamefont {Ulbricht}},
  \bibinfo {author} {\bibfnamefont {K.}~\bibnamefont {Tielrooij}}, \bibinfo
  {author} {\bibfnamefont {A.}~\bibnamefont {Osherov}}, \bibinfo {author}
  {\bibfnamefont {Y.}~\bibnamefont {Golan}}, \bibinfo {author} {\bibfnamefont
  {C.}~\bibnamefont {Delerue}}, \bibinfo {author} {\bibfnamefont
  {G.}~\bibnamefont {Allan}}, \ and\ \bibinfo {author} {\bibfnamefont
  {M.}~\bibnamefont {Bonn}},\ }\bibfield  {title} {\enquote {\bibinfo {title} {Assessment of carrier-multiplication efficiency in bulk PbSe and PbS},}\ }\href
  {https://www.nature.com/articles/nphys1393} {\bibfield  {journal} {\bibinfo
  {journal} {Nat. Phys}\ }\textbf {\bibinfo {volume} {5}},\ \bibinfo {pages}
  {811} (\bibinfo {year} {2009})}\BibitemShut {NoStop}%
\bibitem [{\citenamefont {G{\'e}linas}\ \emph {et~al.}(2013)\citenamefont
  {G{\'e}linas}, \citenamefont {Rao}, \citenamefont {Kumar}, \citenamefont
  {Smith}, \citenamefont {Chin}, \citenamefont {Clark}, \citenamefont {van~der
  Poll}, \citenamefont {Bazan},\ and\ \citenamefont
  {Friend}}]{gelinas2013ultrafast}%
  \BibitemOpen
  \bibfield  {author} {\bibinfo {author} {\bibfnamefont {S.}~\bibnamefont
  {G{\'e}linas}}, \bibinfo {author} {\bibfnamefont {A.}~\bibnamefont {Rao}},
  \bibinfo {author} {\bibfnamefont {A.}~\bibnamefont {Kumar}}, \bibinfo
  {author} {\bibfnamefont {S.~L.}\ \bibnamefont {Smith}}, \bibinfo {author}
  {\bibfnamefont {A.~W.}\ \bibnamefont {Chin}}, \bibinfo {author}
  {\bibfnamefont {J.}~\bibnamefont {Clark}}, \bibinfo {author} {\bibfnamefont
  {T.~S.}\ \bibnamefont {van~der Poll}}, \bibinfo {author} {\bibfnamefont
  {G.~C.}\ \bibnamefont {Bazan}}, \ and\ \bibinfo {author} {\bibfnamefont
  {R.~H.}\ \bibnamefont {Friend}},\ }\bibfield  {title} {\enquote {\bibinfo {title} {Ultrafast long-Range charge separation in organic semiconductor photovoltaic diodes},}\ }\href {\doibase 10.1126/science.1246249}
  {\bibfield  {journal} {\bibinfo  {journal} {Science}\ \textbf {\bibinfo {volume} {343}},\ \bibinfo {pages}
  {512}} (\bibinfo {year} {2014})}\BibitemShut {NoStop}%
\bibitem [{\citenamefont {Jailaubekov}\ \emph {et~al.}(2013)\citenamefont
  {Jailaubekov}, \citenamefont {Willard}, \citenamefont {Tritsch},
  \citenamefont {Chan}, \citenamefont {Sai}, \citenamefont {Gearba},
  \citenamefont {Kaake}, \citenamefont {Williams}, \citenamefont {Leung},
  \citenamefont {Rossky},\ and\ \citenamefont {Zhu}}]{jailaubekov2013hot}%
  \BibitemOpen
  \bibfield  {author} {\bibinfo {author} {\bibfnamefont {A.~E.}\ \bibnamefont
  {Jailaubekov}}, \bibinfo {author} {\bibfnamefont {A.~P.}\ \bibnamefont
  {Willard}}, \bibinfo {author} {\bibfnamefont {J.~R.}\ \bibnamefont
  {Tritsch}}, \bibinfo {author} {\bibfnamefont {W.-L.}\ \bibnamefont {Chan}},
  \bibinfo {author} {\bibfnamefont {N.}~\bibnamefont {Sai}}, \bibinfo {author}
  {\bibfnamefont {R.}~\bibnamefont {Gearba}}, \bibinfo {author} {\bibfnamefont
  {L.~G.}\ \bibnamefont {Kaake}}, \bibinfo {author} {\bibfnamefont {K.~J.}\
  \bibnamefont {Williams}}, \bibinfo {author} {\bibfnamefont {K.}~\bibnamefont
  {Leung}}, \bibinfo {author} {\bibfnamefont {P.~J.}\ \bibnamefont {Rossky}}, \
  and\ \bibinfo {author} {\bibfnamefont {X.-Y.}\ \bibnamefont {Zhu}},\ }\bibfield  {title} {\enquote {\bibinfo {title} {Hot charge-transfer excitons set the time limit for charge separation at donor/acceptor interfaces in organic photovoltaics},}\ }\href
  {https://www.nature.com/articles/nmat3500} {\bibfield  {journal} {\bibinfo
  {journal} {Nat. Mater}\ }\textbf {\bibinfo {volume} {12}},\ \bibinfo {pages}
  {66} (\bibinfo {year} {2013})}\BibitemShut {NoStop}%
\bibitem [{\citenamefont {Jakowetz}\ \emph {et~al.}(2017)\citenamefont
  {Jakowetz}, \citenamefont {B{\"o}hm}, \citenamefont {Sadhanala},
  \citenamefont {Huettner}, \citenamefont {Rao},\ and\ \citenamefont
  {Friend}}]{jakowetz2017visualizing}%
  \BibitemOpen
  \bibfield  {author} {\bibinfo {author} {\bibfnamefont {A.~C.}\ \bibnamefont
  {Jakowetz}}, \bibinfo {author} {\bibfnamefont {M.~L.}\ \bibnamefont
  {B{\"o}hm}}, \bibinfo {author} {\bibfnamefont {A.}~\bibnamefont {Sadhanala}},
  \bibinfo {author} {\bibfnamefont {S.}~\bibnamefont {Huettner}}, \bibinfo
  {author} {\bibfnamefont {A.}~\bibnamefont {Rao}}, \ and\ \bibinfo {author}
  {\bibfnamefont {R.~H.}\ \bibnamefont {Friend}},\ }\bibfield  {title} {\enquote {\bibinfo {title} {Visualizing excitations at buried heterojunctions in organic semiconductor blends},}\ }\href
  {https://www.nature.com/articles/nmat4865} {\bibfield  {journal} {\bibinfo
  {journal} {Nat. Mater}\ }\textbf {\bibinfo {volume} {16}},\ \bibinfo {pages}
  {551} (\bibinfo {year} {2017})}\BibitemShut {NoStop}%
\bibitem [{\citenamefont {Blancon}\ \emph {et~al.}(2017)\citenamefont
  {Blancon}, \citenamefont {Tsai}, \citenamefont {Nie}, \citenamefont
  {Stoumpos}, \citenamefont {Pedesseau}, \citenamefont {Katan}, \citenamefont
  {Kepenekian}, \citenamefont {Soe}, \citenamefont {Appavoo}, \citenamefont
  {Sfeir}, \citenamefont {Tretiak}, \citenamefont {Ajayan}, \citenamefont
  {Kanatzidis}, \citenamefont {Even}, \citenamefont {Crochet},\ and\
  \citenamefont {Mohite}}]{blancon2017extremely}%
  \BibitemOpen
  \bibfield  {author} {\bibinfo {author} {\bibfnamefont {J.-C.}\ \bibnamefont
  {Blancon}}, \bibinfo {author} {\bibfnamefont {H.}~\bibnamefont {Tsai}},
  \bibinfo {author} {\bibfnamefont {W.}~\bibnamefont {Nie}}, \bibinfo {author}
  {\bibfnamefont {C.~C.}\ \bibnamefont {Stoumpos}}, \bibinfo {author}
  {\bibfnamefont {L.}~\bibnamefont {Pedesseau}}, \bibinfo {author}
  {\bibfnamefont {C.}~\bibnamefont {Katan}}, \bibinfo {author} {\bibfnamefont
  {M.}~\bibnamefont {Kepenekian}}, \bibinfo {author} {\bibfnamefont {C.~M.~M.}\
  \bibnamefont {Soe}}, \bibinfo {author} {\bibfnamefont {K.}~\bibnamefont
  {Appavoo}}, \bibinfo {author} {\bibfnamefont {M.~Y.}\ \bibnamefont {Sfeir}},
  \bibinfo {author} {\bibfnamefont {S.}~\bibnamefont {Tretiak}}, \bibinfo
  {author} {\bibfnamefont {P.~M.}\ \bibnamefont {Ajayan}}, \bibinfo {author}
  {\bibfnamefont {M.~G.}\ \bibnamefont {Kanatzidis}}, \bibinfo {author}
  {\bibfnamefont {J.}~\bibnamefont {Even}}, \bibinfo {author} {\bibfnamefont
  {J.~J.}\ \bibnamefont {Crochet}}, \ and\ \bibinfo {author} {\bibfnamefont
  {A.~D.}\ \bibnamefont {Mohite}},\ }\bibfield  {title} {\enquote {\bibinfo {title} {Extremely efficient internal exciton dissociation through edge states in layered 2D perovskites},}\ }\href
  {http://science.sciencemag.org/content/early/2017/03/08/science.aal4211}
  {\bibfield  {journal} {\bibinfo  {journal} {Science}\ }\textbf {\bibinfo
  {volume} {355}},\ \bibinfo {pages} {1288} (\bibinfo {year}
  {2017})}\BibitemShut {NoStop}%
\bibitem [{\citenamefont {Sablon}\ \emph {et~al.}(2011)\citenamefont {Sablon},
  \citenamefont {Little}, \citenamefont {Mitin}, \citenamefont {Sergeev},
  \citenamefont {Vagidov},\ and\ \citenamefont
  {Reinhardt}}]{kimberly2011strong}%
  \BibitemOpen
  \bibfield  {author} {\bibinfo {author} {\bibfnamefont {K.~A.}\ \bibnamefont
  {Sablon}}, \bibinfo {author} {\bibfnamefont {J.~W.}\ \bibnamefont {Little}},
  \bibinfo {author} {\bibfnamefont {V.}~\bibnamefont {Mitin}}, \bibinfo
  {author} {\bibfnamefont {A.}~\bibnamefont {Sergeev}}, \bibinfo {author}
  {\bibfnamefont {N.}~\bibnamefont {Vagidov}}, \ and\ \bibinfo {author}
  {\bibfnamefont {K.}~\bibnamefont {Reinhardt}},\ }\bibfield  {title} {\enquote {\bibinfo {title} {Strong enhancement of solar cell efficiency due to quantum dots with built-in charge},}\ }\href {\doibase
  10.1021/nl200543v} {\bibfield  {journal} {\bibinfo  {journal} {Nano Lett}\
  }\textbf {\bibinfo {volume} {11}},\ \bibinfo {pages} {2311} (\bibinfo {year}
  {2011})}\BibitemShut {NoStop}%
\bibitem [{\citenamefont {Zrenner}\ \emph {et~al.}(2002)\citenamefont
  {Zrenner}, \citenamefont {Beham}, \citenamefont {Stufler}, \citenamefont
  {Findeis}, \citenamefont {Bichler},\ and\ \citenamefont
  {Abstreiter}}]{zrenner2002coherent}%
  \BibitemOpen
  \bibfield  {author} {\bibinfo {author} {\bibfnamefont {A.}~\bibnamefont
  {Zrenner}}, \bibinfo {author} {\bibfnamefont {E.}~\bibnamefont {Beham}},
  \bibinfo {author} {\bibfnamefont {S.}~\bibnamefont {Stufler}}, \bibinfo
  {author} {\bibfnamefont {F.}~\bibnamefont {Findeis}}, \bibinfo {author}
  {\bibfnamefont {M.}~\bibnamefont {Bichler}}, \ and\ \bibinfo {author}
  {\bibfnamefont {G.}~\bibnamefont {Abstreiter}},\ }\bibfield  {title} {\enquote {\bibinfo {title} {Coherent properties of a two-level system based on a quantum-dot photodiode},}\ }\href
  {http://dx.doi.org/10.1038/nature00912} {\bibfield  {journal} {\bibinfo
  {journal} {Nature (London)}\ }\textbf {\bibinfo {volume} {418}},\ \bibinfo
  {pages} {612} (\bibinfo {year} {2002})}\BibitemShut {NoStop}%
\bibitem [{\citenamefont {Ramsay}\ \emph {et~al.}(2008)\citenamefont {Ramsay},
  \citenamefont {Boyle}, \citenamefont {Kolodka}, \citenamefont {Oliveira},
  \citenamefont {Skiba-Szymanska}, \citenamefont {Liu}, \citenamefont
  {Hopkinson}, \citenamefont {Fox},\ and\ \citenamefont
  {Skolnick}}]{Ramsay2008Fast}%
  \BibitemOpen
  \bibfield  {author} {\bibinfo {author} {\bibfnamefont {A.~J.}\ \bibnamefont
  {Ramsay}}, \bibinfo {author} {\bibfnamefont {S.~J.}\ \bibnamefont {Boyle}},
  \bibinfo {author} {\bibfnamefont {R.~S.}\ \bibnamefont {Kolodka}}, \bibinfo
  {author} {\bibfnamefont {J.~B.~B.}\ \bibnamefont {Oliveira}}, \bibinfo
  {author} {\bibfnamefont {J.}~\bibnamefont {Skiba-Szymanska}}, \bibinfo
  {author} {\bibfnamefont {H.~Y.}\ \bibnamefont {Liu}}, \bibinfo {author}
  {\bibfnamefont {M.}~\bibnamefont {Hopkinson}}, \bibinfo {author}
  {\bibfnamefont {A.~M.}\ \bibnamefont {Fox}}, \ and\ \bibinfo {author}
  {\bibfnamefont {M.~S.}\ \bibnamefont {Skolnick}},\ }\bibfield  {title} {\enquote {\bibinfo {title} {Fast Optical Preparation, Control, and Readout of a Single Quantum Dot Spin},}\ }\href {\doibase
  10.1103/PhysRevLett.100.197401} {\bibfield  {journal} {\bibinfo  {journal}
  {Phys. Rev. Lett.}\ }\textbf {\bibinfo {volume} {100}},\ \bibinfo {pages}
  {197401} (\bibinfo {year} {2008})}\BibitemShut {NoStop}%
\bibitem [{\citenamefont {Mar}\ \emph {et~al.}(2014)\citenamefont {Mar},
  \citenamefont {Baumberg}, \citenamefont {Xu}, \citenamefont {Irvine},\ and\
  \citenamefont {Williams}}]{mar2014ultrafast}%
  \BibitemOpen
  \bibfield  {author} {\bibinfo {author} {\bibfnamefont {J.~D.}\ \bibnamefont
  {Mar}}, \bibinfo {author} {\bibfnamefont {J.~J.}\ \bibnamefont {Baumberg}},
  \bibinfo {author} {\bibfnamefont {X.}~\bibnamefont {Xu}}, \bibinfo {author}
  {\bibfnamefont {A.~C.}\ \bibnamefont {Irvine}}, \ and\ \bibinfo {author}
  {\bibfnamefont {D.~A.}\ \bibnamefont {Williams}},\ }\bibfield  {title} {\enquote {\bibinfo {title} {Ultrafast high-fidelity initialization of a quantum-dot spin qubit without magnetic fields},}\ }\href {\doibase
  10.1103/PhysRevB.90.241303} {\bibfield  {journal} {\bibinfo  {journal} {Phys.
  Rev. B}\ }\textbf {\bibinfo {volume} {90}},\ \bibinfo {pages} {241303}
  (\bibinfo {year} {2014})}\BibitemShut {NoStop}%
\bibitem [{\citenamefont {Nozawa}\ \emph {et~al.}(2015)\citenamefont {Nozawa},
  \citenamefont {Takagi}, \citenamefont {Watanabe},\ and\ \citenamefont
  {Arakawa}}]{tomohiro2015direct}%
  \BibitemOpen
  \bibfield  {author} {\bibinfo {author} {\bibfnamefont {T.}~\bibnamefont
  {Nozawa}}, \bibinfo {author} {\bibfnamefont {H.}~\bibnamefont {Takagi}},
  \bibinfo {author} {\bibfnamefont {K.}~\bibnamefont {Watanabe}}, \ and\
  \bibinfo {author} {\bibfnamefont {Y.}~\bibnamefont {Arakawa}},\ }\bibfield  {title} {\enquote {\bibinfo {title} {Direct observation of two-step photon absorption in an InAs/GaAs single quantum dot for the operation of intermediate-band solar cells},}\ }\href
  {\doibase 10.1021/acs.nanolett.5b00947} {\bibfield  {journal} {\bibinfo
  {journal} {Nano Lett}\ }\textbf {\bibinfo {volume} {15}},\ \bibinfo {pages}
  {4483} (\bibinfo {year} {2015})}\BibitemShut {NoStop}%
\bibitem [{\citenamefont {Bayer}\ \emph {et~al.}(2002)\citenamefont {Bayer},
  \citenamefont {Ortner}, \citenamefont {Stern}, \citenamefont {Kuther},
  \citenamefont {Gorbunov}, \citenamefont {Forchel}, \citenamefont {Hawrylak},
  \citenamefont {Fafard}, \citenamefont {Hinzer}, \citenamefont {Reinecke},
  \citenamefont {Walck}, \citenamefont {Reithmaier}, \citenamefont {Klopf},\
  and\ \citenamefont {Sch\"afer}}]{bayer2002fine}%
  \BibitemOpen
  \bibfield  {author} {\bibinfo {author} {\bibfnamefont {M.}~\bibnamefont
  {Bayer}}, \bibinfo {author} {\bibfnamefont {G.}~\bibnamefont {Ortner}},
  \bibinfo {author} {\bibfnamefont {O.}~\bibnamefont {Stern}}, \bibinfo
  {author} {\bibfnamefont {A.}~\bibnamefont {Kuther}}, \bibinfo {author}
  {\bibfnamefont {A.~A.}\ \bibnamefont {Gorbunov}}, \bibinfo {author}
  {\bibfnamefont {A.}~\bibnamefont {Forchel}}, \bibinfo {author} {\bibfnamefont
  {P.}~\bibnamefont {Hawrylak}}, \bibinfo {author} {\bibfnamefont
  {S.}~\bibnamefont {Fafard}}, \bibinfo {author} {\bibfnamefont
  {K.}~\bibnamefont {Hinzer}}, \bibinfo {author} {\bibfnamefont {T.~L.}\
  \bibnamefont {Reinecke}}, \bibinfo {author} {\bibfnamefont {S.~N.}\
  \bibnamefont {Walck}}, \bibinfo {author} {\bibfnamefont {J.~P.}\ \bibnamefont
  {Reithmaier}}, \bibinfo {author} {\bibfnamefont {F.}~\bibnamefont {Klopf}}, \
  and\ \bibinfo {author} {\bibfnamefont {F.}~\bibnamefont {Sch\"afer}},\ }\bibfield  {title} {\enquote {\bibinfo {title} {Fine structure of neutral and charged excitons in self-assembled In(Ga)As/(Al)GaAs quantum dots},}\ }\href
  {\doibase 10.1103/PhysRevB.65.195315} {\bibfield  {journal} {\bibinfo
  {journal} {Phys. Rev. B}\ }\textbf {\bibinfo {volume} {65}},\ \bibinfo
  {pages} {195315} (\bibinfo {year} {2002})}\BibitemShut {NoStop}%
\bibitem [{\citenamefont {Mar}\ \emph {et~al.}(2011)\citenamefont {Mar},
  \citenamefont {Xu}, \citenamefont {Baumberg}, \citenamefont {Brossard},
  \citenamefont {Irvine}, \citenamefont {Stanley},\ and\ \citenamefont
  {Williams}}]{Mar2011bias}%
  \BibitemOpen
  \bibfield  {author} {\bibinfo {author} {\bibfnamefont {J.~D.}\ \bibnamefont
  {Mar}}, \bibinfo {author} {\bibfnamefont {X.~L.}\ \bibnamefont {Xu}},
  \bibinfo {author} {\bibfnamefont {J.~J.}\ \bibnamefont {Baumberg}}, \bibinfo
  {author} {\bibfnamefont {F.~S.~F.}\ \bibnamefont {Brossard}}, \bibinfo
  {author} {\bibfnamefont {A.~C.}\ \bibnamefont {Irvine}}, \bibinfo {author}
  {\bibfnamefont {C.}~\bibnamefont {Stanley}}, \ and\ \bibinfo {author}
  {\bibfnamefont {D.~A.}\ \bibnamefont {Williams}},\ }\bibfield  {title} {\enquote {\bibinfo {title} {Bias-controlled single-electron charging of a self-assembled quantum dot in a two-dimensional-electron-gas-based n-i-Schottky diode},}\ }\href {\doibase
  10.1103/PhysRevB.83.075306} {\bibfield  {journal} {\bibinfo  {journal} {Phys.
  Rev. B}\ }\textbf {\bibinfo {volume} {83}},\ \bibinfo {pages} {075306}
  (\bibinfo {year} {2011})}\BibitemShut {NoStop}%
\bibitem [{\citenamefont {Peng}\ \emph {et~al.}(2017)\citenamefont {Peng},
  \citenamefont {Wu}, \citenamefont {Tang}, \citenamefont {Song}, \citenamefont
  {Qian}, \citenamefont {Sun}, \citenamefont {Xiao}, \citenamefont {Wang},
  \citenamefont {Ali}, \citenamefont {Williams},\ and\ \citenamefont
  {Xu}}]{peng2017probing}%
  \BibitemOpen
  \bibfield  {author} {\bibinfo {author} {\bibfnamefont {K.}~\bibnamefont
  {Peng}}, \bibinfo {author} {\bibfnamefont {S.}~\bibnamefont {Wu}}, \bibinfo
  {author} {\bibfnamefont {J.}~\bibnamefont {Tang}}, \bibinfo {author}
  {\bibfnamefont {F.}~\bibnamefont {Song}}, \bibinfo {author} {\bibfnamefont
  {C.}~\bibnamefont {Qian}}, \bibinfo {author} {\bibfnamefont {S.}~\bibnamefont
  {Sun}}, \bibinfo {author} {\bibfnamefont {S.}~\bibnamefont {Xiao}}, \bibinfo
  {author} {\bibfnamefont {M.}~\bibnamefont {Wang}}, \bibinfo {author}
  {\bibfnamefont {H.}~\bibnamefont {Ali}}, \bibinfo {author} {\bibfnamefont
  {D.~A.}\ \bibnamefont {Williams}}, \ and\ \bibinfo {author} {\bibfnamefont
  {X.}~\bibnamefont {Xu}},\ }\bibfield  {title} {\enquote {\bibinfo {title} {Probing the Dark-Exciton States of a Single Quantum Dot Using Photocurrent Spectroscopy in a Magnetic Field},}\ }\href {\doibase 10.1103/PhysRevApplied.8.064018}
  {\bibfield  {journal} {\bibinfo  {journal} {Phys. Rev. Applied}\ }\textbf
  {\bibinfo {volume} {8}},\ \bibinfo {pages} {064018} (\bibinfo {year}
  {2017})}\BibitemShut {NoStop}%
\bibitem [{\citenamefont {Mar}\ \emph {et~al.}(2013)\citenamefont {Mar},
  \citenamefont {Baumberg}, \citenamefont {Xu}, \citenamefont {Irvine},
  \citenamefont {Stanley},\ and\ \citenamefont {Williams}}]{mar2013high}%
  \BibitemOpen
  \bibfield  {author} {\bibinfo {author} {\bibfnamefont {J.~D.}\ \bibnamefont
  {Mar}}, \bibinfo {author} {\bibfnamefont {J.~J.}\ \bibnamefont {Baumberg}},
  \bibinfo {author} {\bibfnamefont {X.~L.}\ \bibnamefont {Xu}}, \bibinfo
  {author} {\bibfnamefont {A.~C.}\ \bibnamefont {Irvine}}, \bibinfo {author}
  {\bibfnamefont {C.~R.}\ \bibnamefont {Stanley}}, \ and\ \bibinfo {author}
  {\bibfnamefont {D.~A.}\ \bibnamefont {Williams}},\ }\bibfield  {title} {\enquote {\bibinfo {title} {High-resolution photocurrent spectroscopy of the positive trion state in a single quantum dot},}\ }\href {\doibase
  10.1103/PhysRevB.87.155315} {\bibfield  {journal} {\bibinfo  {journal} {Phys.
  Rev. B}\ }\textbf {\bibinfo {volume} {87}},\ \bibinfo {pages} {155315}
  (\bibinfo {year} {2013})}\BibitemShut {NoStop}%
\bibitem [{\citenamefont {Mar}\ \emph {et~al.}(2011{\natexlab{a}})\citenamefont
  {Mar}, \citenamefont {Xu}, \citenamefont {Baumberg}, \citenamefont {Irvine},
  \citenamefont {Stanley},\ and\ \citenamefont {Williams}}]{mar2011voltage}%
  \BibitemOpen
  \bibfield  {author} {\bibinfo {author} {\bibfnamefont {J.~D.}\ \bibnamefont
  {Mar}}, \bibinfo {author} {\bibfnamefont {X.~L.}\ \bibnamefont {Xu}},
  \bibinfo {author} {\bibfnamefont {J.~J.}\ \bibnamefont {Baumberg}}, \bibinfo
  {author} {\bibfnamefont {A.~C.}\ \bibnamefont {Irvine}}, \bibinfo {author}
  {\bibfnamefont {C.}~\bibnamefont {Stanley}}, \ and\ \bibinfo {author}
  {\bibfnamefont {D.~A.}\ \bibnamefont {Williams}},\ }\bibfield  {title} {\enquote {\bibinfo {title} {Voltage-controlled electron tunneling from a single self-assembled quantum dot embedded in a two-dimensional-electron-gas-based photovoltaic cell},}\ }\href {\doibase
  10.1063/1.3633216} {\bibfield  {journal} {\bibinfo  {journal} {J. Appl.
  Phys.}\ }\textbf {\bibinfo {volume} {110}},\ \bibinfo {pages} {053110}
  (\bibinfo {year} {2011}{\natexlab{a}})}\BibitemShut {NoStop}%
\bibitem [{\citenamefont {Mar}\ \emph {et~al.}(2011{\natexlab{b}})\citenamefont
  {Mar}, \citenamefont {Xu}, \citenamefont {Baumberg}, \citenamefont {Irvine},
  \citenamefont {Stanley},\ and\ \citenamefont
  {Williams}}]{mar2011electrically}%
  \BibitemOpen
  \bibfield  {author} {\bibinfo {author} {\bibfnamefont {J.~D.}\ \bibnamefont
  {Mar}}, \bibinfo {author} {\bibfnamefont {X.~L.}\ \bibnamefont {Xu}},
  \bibinfo {author} {\bibfnamefont {J.~J.}\ \bibnamefont {Baumberg}}, \bibinfo
  {author} {\bibfnamefont {A.~C.}\ \bibnamefont {Irvine}}, \bibinfo {author}
  {\bibfnamefont {C.}~\bibnamefont {Stanley}}, \ and\ \bibinfo {author}
  {\bibfnamefont {D.~A.}\ \bibnamefont {Williams}},\ }\bibfield  {title} {\enquote {\bibinfo {title} {Electrically tunable hole tunnelling from a single self-assembled quantum dot embedded in an \emph{n-i-}Schottky photovoltaic cell},}\ }\href {\doibase
  10.1063/1.3614418} {\bibfield  {journal} {\bibinfo  {journal} {Appl. Phys.
  Lett}\ }\textbf {\bibinfo {volume} {99}},\ \bibinfo {pages} {031102}
  (\bibinfo {year} {2011}{\natexlab{b}})}\BibitemShut {NoStop}%
\bibitem [{\citenamefont {Nguyen}\ \emph {et~al.}(2012)\citenamefont {Nguyen},
  \citenamefont {Sallen}, \citenamefont {Voisin}, \citenamefont {Roussignol},
  \citenamefont {Diederichs},\ and\ \citenamefont
  {Cassabois}}]{nguyen2012optically}%
  \BibitemOpen
  \bibfield  {author} {\bibinfo {author} {\bibfnamefont {H.~S.}\ \bibnamefont
  {Nguyen}}, \bibinfo {author} {\bibfnamefont {G.}~\bibnamefont {Sallen}},
  \bibinfo {author} {\bibfnamefont {C.}~\bibnamefont {Voisin}}, \bibinfo
  {author} {\bibfnamefont {P.}~\bibnamefont {Roussignol}}, \bibinfo {author}
  {\bibfnamefont {C.}~\bibnamefont {Diederichs}}, \ and\ \bibinfo {author}
  {\bibfnamefont {G.}~\bibnamefont {Cassabois}},\ }\bibfield  {title} {\enquote {\bibinfo {title} {Optically Gated Resonant Emission of Single Quantum Dots},}\ }\href {\doibase
  10.1103/PhysRevLett.108.057401} {\bibfield  {journal} {\bibinfo  {journal}
  {Phys. Rev. Lett.}\ }\textbf {\bibinfo {volume} {108}},\ \bibinfo {pages}
  {057401} (\bibinfo {year} {2012})}\BibitemShut {NoStop}%
\bibitem [{\citenamefont {Bennett}\ \emph {et~al.}(2016)\citenamefont
  {Bennett}, \citenamefont {Lee}, \citenamefont {Ellis}, \citenamefont {Meany},
  \citenamefont {Murray}, \citenamefont {Floether}, \citenamefont {Griffths},
  \citenamefont {Farrer}, \citenamefont {Ritchie},\ and\ \citenamefont
  {Shields}}]{bennett2016cavity}%
  \BibitemOpen
  \bibfield  {author} {\bibinfo {author} {\bibfnamefont {A.~J.}\ \bibnamefont
  {Bennett}}, \bibinfo {author} {\bibfnamefont {J.~P.}\ \bibnamefont {Lee}},
  \bibinfo {author} {\bibfnamefont {D.~J.~P.}\ \bibnamefont {Ellis}}, \bibinfo
  {author} {\bibfnamefont {T.}~\bibnamefont {Meany}}, \bibinfo {author}
  {\bibfnamefont {E.}~\bibnamefont {Murray}}, \bibinfo {author} {\bibfnamefont
  {F.~F.}\ \bibnamefont {Floether}}, \bibinfo {author} {\bibfnamefont {J.~P.}\
  \bibnamefont {Griffths}}, \bibinfo {author} {\bibfnamefont {I.}~\bibnamefont
  {Farrer}}, \bibinfo {author} {\bibfnamefont {D.~A.}\ \bibnamefont {Ritchie}},
  \ and\ \bibinfo {author} {\bibfnamefont {A.~J.}\ \bibnamefont {Shields}},\
  }\bibfield  {title} {\enquote {\bibinfo {title} {Cavity-enhanced coherent light scattering from a quantum dot},}\ }\href {\doibase 10.1126/sciadv.1501256} {\bibfield  {journal} {\bibinfo
  {journal} {Sci. Adv.}\ }\textbf {\bibinfo {volume} {2}},\ \bibinfo {pages}
  {e1501256} (\bibinfo {year} {2016})}\BibitemShut {NoStop}%
\bibitem [{\citenamefont {Moody}\ \emph {et~al.}(2016)\citenamefont {Moody},
  \citenamefont {McDonald}, \citenamefont {Feldman}, \citenamefont {Harvey},
  \citenamefont {Mirin},\ and\ \citenamefont
  {Silverman}}]{moody2016electronic}%
  \BibitemOpen
  \bibfield  {author} {\bibinfo {author} {\bibfnamefont {G.}~\bibnamefont
  {Moody}}, \bibinfo {author} {\bibfnamefont {C.}~\bibnamefont {McDonald}},
  \bibinfo {author} {\bibfnamefont {A.}~\bibnamefont {Feldman}}, \bibinfo
  {author} {\bibfnamefont {T.}~\bibnamefont {Harvey}}, \bibinfo {author}
  {\bibfnamefont {R.~P.}\ \bibnamefont {Mirin}}, \ and\ \bibinfo {author}
  {\bibfnamefont {K.~L.}\ \bibnamefont {Silverman}},\ }\bibfield  {title} {\enquote {\bibinfo {title} {Electronic Enhancement of the Exciton Coherence Time in Charged Quantum Dots},}\ }\href {\doibase
  10.1103/PhysRevLett.116.037402} {\bibfield  {journal} {\bibinfo  {journal}
  {Phys. Rev. Lett.}\ }\textbf {\bibinfo {volume} {116}},\ \bibinfo {pages}
  {037402} (\bibinfo {year} {2016})}\BibitemShut {NoStop}%
\bibitem [{\citenamefont {Kurzmann}\ \emph {et~al.}(2016)\citenamefont
  {Kurzmann}, \citenamefont {Ludwig}, \citenamefont {Wieck}, \citenamefont
  {Lorke},\ and\ \citenamefont {Geller}}]{kurzmann2016auger}%
  \BibitemOpen
  \bibfield  {author} {\bibinfo {author} {\bibfnamefont {A.}~\bibnamefont
  {Kurzmann}}, \bibinfo {author} {\bibfnamefont {A.}~\bibnamefont {Ludwig}},
  \bibinfo {author} {\bibfnamefont {A.~D.}\ \bibnamefont {Wieck}}, \bibinfo
  {author} {\bibfnamefont {A.}~\bibnamefont {Lorke}}, \ and\ \bibinfo {author}
  {\bibfnamefont {M.}~\bibnamefont {Geller}},\ }\bibfield  {title} {\enquote {\bibinfo {title} {Auger recombination in self-assembled quantum dots: quenching and broadening of the charged exciton transition},}\ }\href {\doibase
  10.1021/acs.nanolett.6b01082} {\bibfield  {journal} {\bibinfo  {journal}
  {Nano Lett}\ }\textbf {\bibinfo {volume} {16}},\ \bibinfo {pages} {3367}
  (\bibinfo {year} {2016})}\BibitemShut {NoStop}%
\bibitem [{\citenamefont {Beham}\ \emph {et~al.}(2001)\citenamefont {Beham},
  \citenamefont {Zrenner}, \citenamefont {Findeis}, \citenamefont {Bichler},\
  and\ \citenamefont {Abstreiter}}]{beham2001nonlinear}%
  \BibitemOpen
  \bibfield  {author} {\bibinfo {author} {\bibfnamefont {E.}~\bibnamefont
  {Beham}}, \bibinfo {author} {\bibfnamefont {A.}~\bibnamefont {Zrenner}},
  \bibinfo {author} {\bibfnamefont {F.}~\bibnamefont {Findeis}}, \bibinfo
  {author} {\bibfnamefont {M.}~\bibnamefont {Bichler}}, \ and\ \bibinfo
  {author} {\bibfnamefont {G.}~\bibnamefont {Abstreiter}},\ }\bibfield  {title} {\enquote {\bibinfo {title} {Nonlinear ground-state absorption observed in a single quantum dot},}\ }\href {\doibase
  10.1063/1.1411987} {\bibfield  {journal} {\bibinfo  {journal} {Appl. Phys.
  Lett}\ }\textbf {\bibinfo {volume} {79}},\ \bibinfo {pages} {2808} (\bibinfo
  {year} {2001})}\BibitemShut {NoStop}%
\bibitem [{\citenamefont {Heller}\ \emph {et~al.}(1998)\citenamefont {Heller},
  \citenamefont {Bockelmann},\ and\ \citenamefont
  {Abstreiter}}]{heller1998electric}%
  \BibitemOpen
  \bibfield  {author} {\bibinfo {author} {\bibfnamefont {W.}~\bibnamefont
  {Heller}}, \bibinfo {author} {\bibfnamefont {U.}~\bibnamefont {Bockelmann}},
  \ and\ \bibinfo {author} {\bibfnamefont {G.}~\bibnamefont {Abstreiter}},\
  }\bibfield  {title} {\enquote {\bibinfo {title} {Electric-field effects on excitons in quantum dots},}\ }\href {\doibase 10.1103/PhysRevB.57.6270} {\bibfield  {journal} {\bibinfo
  {journal} {Phys. Rev. B}\ }\textbf {\bibinfo {volume} {57}},\ \bibinfo
  {pages} {6270} (\bibinfo {year} {1998})}\BibitemShut {NoStop}%
\bibitem [{\citenamefont {Oulton}\ \emph {et~al.}(2002)\citenamefont {Oulton},
  \citenamefont {Finley}, \citenamefont {Ashmore}, \citenamefont {Gregory},
  \citenamefont {Mowbray}, \citenamefont {Skolnick}, \citenamefont {Steer},
  \citenamefont {Liew}, \citenamefont {Migliorato},\ and\ \citenamefont
  {Cullis}}]{oulton2002manipulation}%
  \BibitemOpen
  \bibfield  {author} {\bibinfo {author} {\bibfnamefont {R.}~\bibnamefont
  {Oulton}}, \bibinfo {author} {\bibfnamefont {J.~J.}\ \bibnamefont {Finley}},
  \bibinfo {author} {\bibfnamefont {A.~D.}\ \bibnamefont {Ashmore}}, \bibinfo
  {author} {\bibfnamefont {I.~S.}\ \bibnamefont {Gregory}}, \bibinfo {author}
  {\bibfnamefont {D.~J.}\ \bibnamefont {Mowbray}}, \bibinfo {author}
  {\bibfnamefont {M.~S.}\ \bibnamefont {Skolnick}}, \bibinfo {author}
  {\bibfnamefont {M.~J.}\ \bibnamefont {Steer}}, \bibinfo {author}
  {\bibfnamefont {S.-L.}\ \bibnamefont {Liew}}, \bibinfo {author}
  {\bibfnamefont {M.~A.}\ \bibnamefont {Migliorato}}, \ and\ \bibinfo {author}
  {\bibfnamefont {A.~J.}\ \bibnamefont {Cullis}},\ }\bibfield  {title} {\enquote {\bibinfo {title} {Manipulation of the homogeneous linewidth of an individual In(Ga)As quantum dot},}\ }\href {\doibase
  10.1103/PhysRevB.66.045313} {\bibfield  {journal} {\bibinfo  {journal} {Phys.
  Rev. B}\ }\textbf {\bibinfo {volume} {66}},\ \bibinfo {pages} {045313}
  (\bibinfo {year} {2002})}\BibitemShut {NoStop}%
\bibitem [{\citenamefont {Cao}\ \emph {et~al.}(2015)\citenamefont {Cao},
  \citenamefont {Tang}, \citenamefont {Gao}, \citenamefont {Sun}, \citenamefont
  {Qiu}, \citenamefont {Zhao}, \citenamefont {He}, \citenamefont {Shi},
  \citenamefont {Gu}, \citenamefont {Williams}, \citenamefont {Sheng},
  \citenamefont {Jin},\ and\ \citenamefont {Xu}}]{cao2015longitudinal}%
  \BibitemOpen
  \bibfield  {author} {\bibinfo {author} {\bibfnamefont {S.}~\bibnamefont
  {Cao}}, \bibinfo {author} {\bibfnamefont {J.}~\bibnamefont {Tang}}, \bibinfo
  {author} {\bibfnamefont {Y.}~\bibnamefont {Gao}}, \bibinfo {author}
  {\bibfnamefont {Y.}~\bibnamefont {Sun}}, \bibinfo {author} {\bibfnamefont
  {K.}~\bibnamefont {Qiu}}, \bibinfo {author} {\bibfnamefont {Y.}~\bibnamefont
  {Zhao}}, \bibinfo {author} {\bibfnamefont {M.}~\bibnamefont {He}}, \bibinfo
  {author} {\bibfnamefont {J.-A.}\ \bibnamefont {Shi}}, \bibinfo {author}
  {\bibfnamefont {L.}~\bibnamefont {Gu}}, \bibinfo {author} {\bibfnamefont
  {D.~A.}\ \bibnamefont {Williams}}, \bibinfo {author} {\bibfnamefont
  {W.}~\bibnamefont {Sheng}}, \bibinfo {author} {\bibfnamefont
  {K.}~\bibnamefont {Jin}}, \ and\ \bibinfo {author} {\bibfnamefont
  {X.}~\bibnamefont {Xu}},\ }\bibfield  {title} {\enquote {\bibinfo {title} {Longitudinal wave function control in single quantum dots with an applied magnetic field},}\ }\href {https://www.nature.com/articles/srep08041}
  {\bibfield  {journal} {\bibinfo  {journal} {Sci. Rep.}\ }\textbf {\bibinfo
  {volume} {5}},\ \bibinfo {pages} {8041} (\bibinfo {year} {2015})}\BibitemShut
  {NoStop}%
\end{thebibliography}
%\providecommand{\noopsort}[1]{}\providecommand{\singleletter}[1]{#1}%

\end{document}